\documentclass[aps,pra,twocolumn,superscriptaddress]{revtex4-2}

\usepackage{graphicx}
\usepackage{float}
\usepackage{appendix}
\usepackage{lipsum}
\usepackage{physics}
\usepackage{bbold}
\usepackage{hyperref}
\usepackage{natbib}
\usepackage{amsmath,mathtools}
\usepackage[normalem]{ulem}

\hypersetup{colorlinks=true,linkcolor=blue,citecolor=blue,urlcolor=blue}

\newcommand{\beq}{\begin{equation}}
\newcommand{\eeq}{\end{equation}}
\newcommand{\beqar}{\begin{eqnarray}}
\newcommand{\eeqar}{\end{eqnarray}}
\newcommand{\bea}{\begin{eqnarray}}
\newcommand{\eea}{\end{eqnarray}}
\newcommand{\bcen}{\begin{center}}
\newcommand{\ecen}{\end{center}}

\begin{document}
\title{
Theory and Experimental Demonstration of Quantum Invariant Filtering
}

\author{Loris Maria Cangemi}
\thanks{These authors contributed equally to this work. \\Present address: Dept. of Electrical Engineering and Information Technology, Università degli Studi di Napoli Federico II, via Claudio 21, Napoli, 80125, Italy.}
\affiliation{Department of Chemistry, Bar-Ilan University, Ramat-Gan 52900, Israel}
\affiliation{Institute of Nanotechnology and Advanced Materials, Bar-Ilan University, Ramat-Gan 52900, Israel}
\affiliation{Center for Quantum Entanglement Science and Technology, Bar-Ilan University, Ramat-Gan 52900, Israel}

\author{Yoav Woldiger}
\thanks{These authors contributed equally to this work.}
\affiliation{Institute of Nanotechnology and Advanced Materials, Bar-Ilan University, Ramat-Gan 52900, Israel}
\affiliation{Department of Physics, Bar-Ilan University, Ramat-Gan 52900, Israel}

\author{Amikam Levy}
\email{amikam.levy@biu.ac.il}
\affiliation{Department of Chemistry, Bar-Ilan University, Ramat-Gan 52900, Israel}
\affiliation{Institute of Nanotechnology and Advanced Materials, Bar-Ilan University, Ramat-Gan 52900, Israel}
\affiliation{Center for Quantum Entanglement Science and Technology, Bar-Ilan University, Ramat-Gan 52900, Israel}

\author{Assaf Hamo}
\email{assaf.hamo@biu.ac.il}
\affiliation{Department of Physics, Bar-Ilan University, Ramat-Gan 52900, Israel}
\affiliation{Institute of Nanotechnology and Advanced Materials, Bar-Ilan University, Ramat-Gan 52900, Israel}

\begin{abstract}
Quantum control protocols are typically devised in the time domain, leaving their spectral behavior to emerge only a posteriori. Here, we invert this paradigm. Starting from a target frequency-domain filter, we employ the dynamical-invariant framework to derive the continuous driving fields that enact the chosen spectral response on a qubit.
This approach, Quantum Invariant Filtering (QIF), maps arbitrary finite-impulse responses, including multi-band and phase-sensitive profiles, into experimentally feasible Hamiltonian modulations.
Implemented on a single nitrogen-vacancy center in diamond, the method realizes the prescribed passbands with high fidelity, suppresses noise, and preserves coherence for milliseconds, two orders of magnitude longer than Carr-Purcell-Meiboom-Gill sequences, while remaining robust to $ 50\%$ drive-amplitude errors.
Our results establish QIF as a broadly applicable framework for enhanced quantum control and sensing across diverse physical platforms, including superconducting qubits, trapped ions, and nuclear magnetic resonance systems.
\end{abstract}

\maketitle

\section*{Introduction}

Quantum coherence underpins the transformative potential of emerging quantum technologies, enabling advancements in computation, communication, and sensing~\cite{degen2017quantum,boss2017quantum,schmitt2017submillihertz,glenn2018high,sagi2010process,bluhm2011dephasing,bar2013solid}. However, maintaining coherence in quantum systems remains a fundamental challenge, primarily due to unavoidable interactions with complex, uncontrolled environmental noise. These interactions lead to decoherence, degrading quantum information and imposing stringent limitations on practical implementations. As a result, developing effective strategies for suppressing noise~\cite{lidar1998decoherence,gordon2008optimal,koch2016controlling,suter2016colloquium,levy2018noise,kuprov2023optimal,cangemi2023control} 
are essential for enabling precise and robust quantum control~\cite{lloyd2001engineering,khaneja2005optimal,doria2011optimal,bason2012high, del2013shortcuts,guery2019shortcuts,rembold2020introduction,koch2022quantum,poggi2024universally} in realistic, noise-prone quantum devices.


Dynamical decoupling (DD) techniques~\cite{viola1998dynamical,viola1999dynamical} have significantly advanced quantum control by mitigating environmental noise through carefully timed sequences of $\pi$ pulses. In particular, protocols such as the Carr-Purcell-Meiboom-Gill (CPMG) protocol~\cite{meiboom1958modified} and related methods have shown broad experimental success across diverse platforms~\cite{biercuk2009optimized,du2009preserving,de2010universal,louzon2025robust}. Such DD strategies, including foundational and optimized pulse sequences~\cite{khodjasteh2005fault,uhrig2007keeping,ryan2010robust,bar2012suppression,salhov2024protecting}, remain cornerstones for protecting quantum coherence. However, these approaches are typically designed in the time domain and only afterward interpreted in terms of their frequency-domain filtering properties, limiting their flexibility, resolution, and sensitivity to amplitude or phase errors.

In this work, we introduce Quantum Invariant Filtering (QIF), a quantum control framework that systematically reverses the conventional design paradigm. Rather than initially formulating control protocols in the time domain and subsequently interpreting their spectral characteristics, QIF begins directly from a desired frequency-domain filter specification, defined explicitly by its spectral response and phase characteristics, and analytically constructs the corresponding time-dependent control Hamiltonian. The method leverages the formalism of dynamical invariants~\cite{lewis69,chen2010fast} and quantum response theory, enabling the precise reverse engineering of continuous driving fields that implement the prescribed spectral properties.

Our approach introduces several key advantages. First, it enables the direct integration of any finite-impulse filter response, including single-band, phase-sensitive, and complex multi-band profiles, into the quantum control field. This includes the design and experimental realization of dual-bandpass filters, which cannot be achieved using conventional DD sequences. Second, the control fields produced by QIF are continuous and smooth, offering improved experimental robustness and easier calibration. Finally, by working within the invariant framework, the method provides a natural way to treat perturbative signals and to tailor the quantum system’s response to weak, structured inputs.

We experimentally demonstrate the QIF protocol using a single nitrogen-vacancy (NV) center in diamond.
On this platform, both the target signal and the dephasing noise couple along the $z$-axis, with the noise exhibiting a predominantly low-frequency spectrum~\cite{paladino20141,kim2015decoherence,romach2015spectroscopy}. We design QIF protocols that selectively transmit the signal frequency while suppressing broadband noise. Compared to standard DD sequences, QIF shows superior resilience to amplitude errors and retains coherence over significantly longer durations. Moreover, we demonstrate that QIF can be used to extract both the amplitude and phase of an applied signal, highlighting its utility for quantum sensing.

\begin{figure*}[htbp!]
\centering
\includegraphics[width=\textwidth]{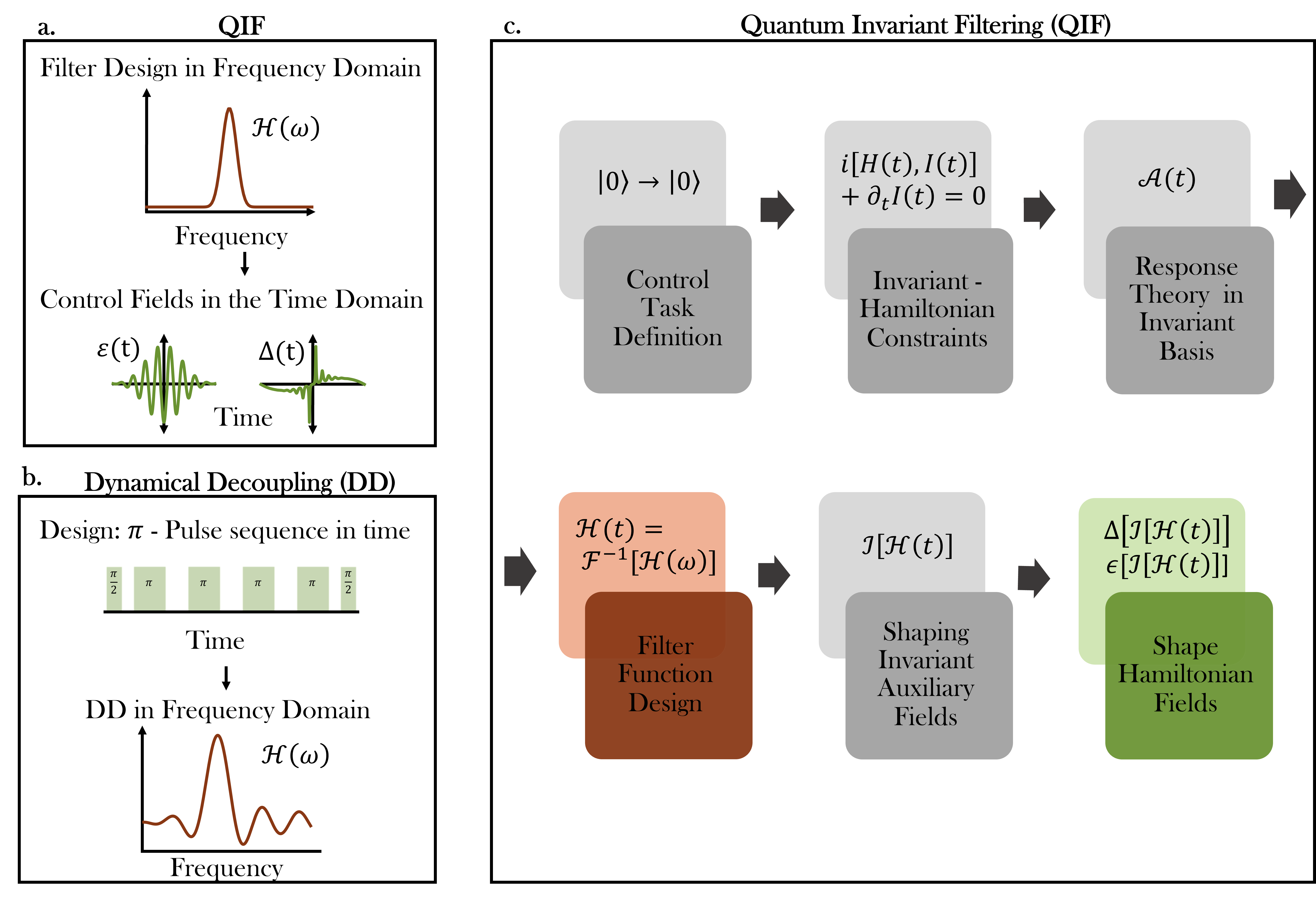}
\caption{\textbf{Schematics of the the QIF protocol:} 
In typical dynamical decoupling approaches \textbf{b.}, control protocols are first designed in the time domain, for example, using sequences of $\pi$ pulses, and their filtering properties are inferred afterward by analyzing the system’s response in the frequency domain, yielding a filter function $\mathcal{H}(\omega)$. In contrast, the Quantum Invariant Filtering (QIF) method \textbf{a.} begins with a desired frequency-domain filter specification and systematically maps it to time-domain control fields, $\Delta(t)$ and $\varepsilon(t)$, enabling targeted and flexible spectral selectivity.
\textbf{c.} Schematic representation of the QIF protocol, which integrates filter design in the frequency domain directly into the construction of a time-dependent control Hamiltonian. This is accomplished through the framework of dynamical invariants, which guide the shaping of auxiliary fields and enable the reverse engineering of control dynamics consistent with the specified spectral response.}
\label{fig:fig1}
\end{figure*}

Altogether, this work establishes a general and scalable framework for embedding frequency-domain filtering into quantum control. The QIF method provides a versatile toolbox for noise suppression and signal detection, with broad applicability to quantum platforms including NV centers, superconducting qubits, trapped ions, and nuclear magnetic resonance.

\section*{Results}

\subsection*{The Quantum Invariant Filtering (QIF) method}
Typical control protocols are first formulated in the time domain and subsequently analyzed for their filtering characteristics through the system’s frequency-domain response, Fig.~\ref{fig:fig1}b. In contrast, the Quantum Invariant Filtering (QIF) approach developed here reverses this process: it starts from a desired filter specification in the frequency domain and directly constructs the corresponding time-dependent control Hamiltonian that realizes the intended spectral response, Fig.~\ref{fig:fig1}a.
This procedure, illustrated in Fig.~\ref{fig:fig1}c, begins by defining the control task through explicit initial and target states. We construct a control field that drives the undisturbed system to the target state with perfect fidelity. To demonstrate the achieved filtering control, we consider the system's response to be a deviation from this ideal final state.
Hence, we select a straightforward scenario, where the system is initialized in the state  $\ket{0}$ (not mandatory), and ideally returns to the same state at the end of the protocol. Deviations from this target explicitly reveal the filter's characteristics, indicating which frequency components and  phasesof the input signal are transmitted or suppressed.

Mapping a frequency-domain filter into a time-dependent control Hamiltonian is inherently challenging, as an analytic solution is essential. To overcome this problem, we adopt the dynamical invariant framework. The dynamical invariant $I(t)$ is a nonunique Hermitian operator satisfying the condition
\begin{equation}
i[H(t), I(t)] + \partial_t I(t) = 0,
\end{equation}
ensuring constant eigenvalues with evolving eigenstates $\{\ket{\phi_k(t)}\}$. Here and throughout
this text, we take $\hbar=1$. By strategically choosing the invariant $I(t)$, particularly selecting its eigenstates to smoothly interpolate between the initial and final target states, we directly solve the invariant condition to determine the corresponding control Hamiltonian $H(t)$ (see Supplementary). This ensures that for an unperturbed system, the state evolves precisely along the predetermined invariant eigenstate.
Working within the invariant basis presents additional advantages, notably expressing the propagator $U(t,t')=\mathcal{T}\exp \left(-i\int_{t'}^{t} H(s)ds \right)$ as
\begin{equation}
U(t,t') = \sum_n e^{i(\varphi_n(t)-\varphi_n(t'))}\ket{\phi_n(t)}\bra{\phi_n(t')},
\end{equation}
with the Lewis-Riesenfeld phase $\varphi_n(t)=\int_{0}^{t} ds \bra{\phi_n(s)} i \partial_s - H(s) \ket{\phi_n(s)}$.
This representation enables a perturbative treatment of external signals directly in terms of the invariant's auxiliary fields, which provide additional degrees of freedom for control. Utilizing these auxiliary fields, we precisely shape the frequency response of the system via known Finite Impulse Response (FIR) filter design methodologies~\cite{oppenheim1997signals} and further phase control by modulation of the impulse response $\mathcal{H}(t)$~\cite{kotler2011single} (see supplementary). Once the auxiliary fields are established, the corresponding Hamiltonian control fields can be immediately determined, enabling finely tailored quantum noise filtering.

For a qubit control Hamiltonian in the rotating wave approximation (RWA) of the form 
$H(t)=\frac{\Delta(t)}{2}\sigma_z -\frac{\varepsilon(t)}{2}\sigma_x$, it is convenient to parametrize the dynamical invariant using the auxiliary fields $\{\alpha(t),\beta(t)\}$ and the Pauli matrices as 
$I(t)= \boldsymbol{{\mathcal{I}}}(t) \cdot \boldsymbol{\sigma}$, 
where
$\boldsymbol{\mathcal{I}}(t)=\frac{1}{2}(-\cos{\alpha(t)},\sin{\alpha(t)}\sin{\beta(t)},\sin{\alpha(t)}\cos{\beta(t)})$. 
The invariant condition then provides a direct relationship between the control fields and the auxiliary fields (Fig.~\ref{fig:fig1}a) given by 

\begin{equation}
\label{eq:fieldeq}
\varepsilon(t)=\dot{\beta}(t) -\frac{\dot{\alpha}(t)}{\tan \alpha(t) \tan \beta(t) }, \qquad \Delta(t)=-\frac{\dot{\alpha}(t)}{\sin \beta(t)}.
\end{equation}

Considering a perturbation along the $z$-axis of the form $V(t) = \frac{\delta}{2} f_{\rm in}(t) \sigma_z$, with an arbitrary time-dependent input signal $f_{\rm in}(t)$, the outcome of $\langle \sigma_z(t_{\rm f}) \rangle$ at the final time $t_{\rm f}$ can be expressed perturbatively in terms of the auxiliary fields $\alpha(t)$ and $\beta(t)$ (see supplementary). In this setting, the dominant noise is assumed to act along the same axis as the input signal, and the filter is designed to suppress noise across all frequencies except for the specific frequency component associated with the target signal.

To satisfy the boundary conditions at the initial and final times, we set $ \alpha(t_b) = \pi$ and $ \beta(t_b) = -\frac{\pi}{2} $ for $ t_b = 0, t_{\rm f} $. To further simplify experimental implementation, we impose $\dot{\alpha}(t) = 0 $ throughout the evolution, which constrains the control field to $ \Delta(t) = 0$, requiring modulation of only $\varepsilon(t)$.
Under this parametrization, we establish a direct connection between the second-order response to an external signal and the auxiliary fields that define the control Hamiltonian. Specifically, the deviation from the ideal outcome can be expressed as
$\langle \sigma_z(t_{\rm f}) \rangle =\langle \sigma_z(t_{\rm f}) \rangle_0 - \frac{1}{2} \mathcal{A}^2(t_{\rm f}) +O(\delta^3),$
where
\begin{equation}
\mathcal{A}(t_{\rm f}) = \delta \int_0^{t_{\rm f}} ds\, f_{\rm in}(s) \cos \beta(s),
\end{equation}
and $ \langle \sigma_z(t_{\rm f}) \rangle_0 = 1$ corresponds to the unperturbed result of the ideal control protocol.

The auxiliary field $\beta(t)$ is constrained by the boundary conditions specified above, but remains undetermined at intermediate times. 
This flexibility enables the integration of a desired filter function $\mathcal{H}(\omega)$ directly into the control Hamiltonian. The corresponding impulse response function is first obtained via inverse discrete Fourier transform, $\mathcal{H}(t) = \mathcal{F}^{-1}[\mathcal{H}(\omega)]$ (see Supplementary). Then, by setting the auxiliary field as $\beta(t) = -\frac{\pi}{2} + \arcsin{\mathcal{H}(t)}$, the system's response to an external input is shaped by the designed filter. In particular, the filtered signal that is manifested in the system's response is given by the convolution of the input signal with the impulse response (see Supplementary):
\begin{equation}
\mathcal{A}(t_{\rm f}) = \delta \int_{-\infty}^{\infty} ds\, f_{\rm in}(s)\, \mathcal{H}(t_{\rm f} - s).
\end{equation}

\begin{figure*}[htbp!]
\includegraphics[width=\textwidth]{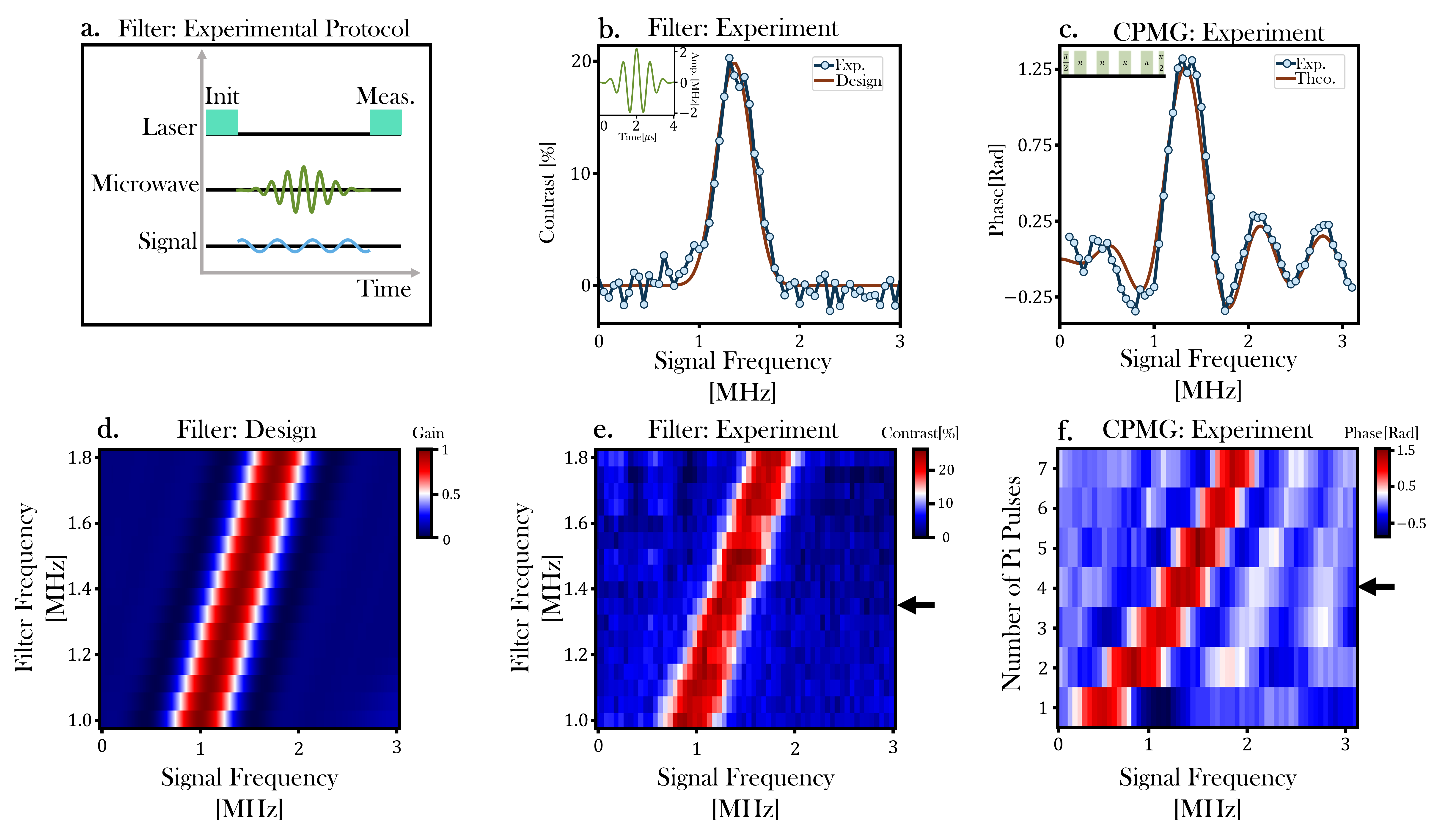}
\caption{\textbf{Quantum Invariant Filter Design: a. Experimental protocol of the Quantum Invariant filter:} The protocol begins with a green laser pulse (top line) initializing the NV center state to the ground state. Immediately following initialization, the control field (middle line) is applied according to the modulation function designed to achieve the desired frequency response. Concurrently, a signal with a specific frequency (bottom line) is applied, matching the phase of the filter unless otherwise stated. \textbf{b. Experimental Quantum Invariant Filter:} The designed filter functions as a bandpass with a central frequency of 1.35 MHz and a bandwidth of 250 kHz. Following the procedure outlined in \textbf{a.}, the modulation function for the microwave field is implemented. A varying frequency signal (0–4 MHz, x-axis) is applied simultaneously with the microwave field. Measurements are taken twice: once with only the filter field applied and once with an additional $\pi$-pulse at the end of the filter field. The y-axis represents the contrast, defined as the difference between these two measurements divided by their average. A distinct response is evident around 1.35 MHz, while other frequencies show no significant response. Experimental results (blue connected dots) represent the contrast of collected photon numbers, while the designed filter response, normalized to the maximum experimental contrast, is indicated by the solid brown line. \textbf{Inset:} Microwave field modulation function. \textbf{c. Carr-Purcell-Meiboom-Gill (CPMG) frequency response comparison:} The filter response of a CPMG pulse sequence with four $\pi$-pulses is presented. Experimental results (blue connected dots) correspond to the measured phase. The theoretical frequency response for the same four-pulse CPMG sequence is illustrated by the solid brown line, clearly displaying frequency ripples. \textbf{Inset:} Illustration of the four-pulse CPMG sequence. \textbf{d. Design of variable filter response:} The frequency responses of multiple bandpass filters are displayed as a colormap. The x-axis corresponds to the signal frequency, and the y-axis represents the central frequency of each designed bandpass filter. The color scale indicates filter response intensity. \textbf{e. Experimental variable filter response:} Following the design outlined in \textbf{d.}, corresponding control fields are generated for each filter frequency. The colormap presents the measured contrast across various filter frequencies (y-axis) and signal frequencies (x-axis), closely matching the designed responses in \textbf{d. f. CPMG variable response comparison:} The frequency response of the CPMG sequence as a function of the number of $\pi$-pulses (with fixed delay times) is shown. The colormap indicates the measured phase, with the x-axis corresponding to the signal frequency and the y-axis indicating the number of $\pi$ pulses within the CPMG sequence.}
\label{fig:fig2}
\end{figure*}

To ensure consistency with the dynamical invariant constraints, the impulse response function $\mathcal{H}(t)$ must vanish at the initial and final times, and be rescaled such that it's amplitude is inside $(-1,1)$.
The corresponding control field is then given by $\varepsilon(t)=-\frac{\partial_t\mathcal{H}(t_{\rm f}-t)}{\sqrt{1-\mathcal{H}(t_{\rm f}-t)^2}}$.  
For experimental convenience, we set $\beta(t) = -\pi/2 + \mathcal{H}(t)$, leading to $\varepsilon(t) = \partial_t \mathcal{H}(t_{\rm f} - t)$ (see Supplementary). This choice ensures that the control field directly implements a filter function with the desired spectral support, encoded by $\mathcal{H}(t)$.


\subsection*{Experiment}

\begin{figure*}[htbp!]
\includegraphics[width=\textwidth]{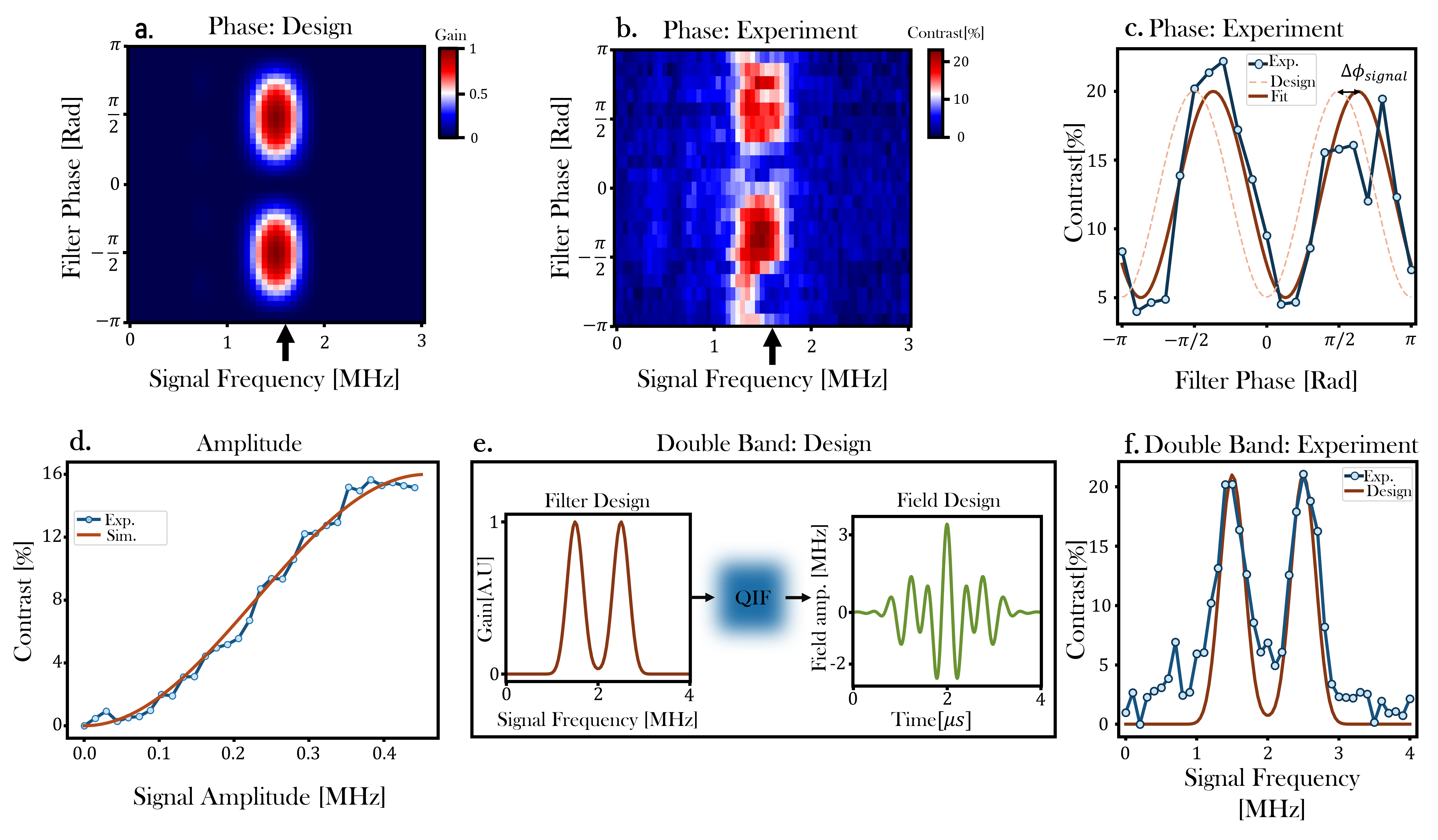}
\caption{\textbf{Quantum Invariant Filter flexibility: }
\textbf{a. Phase dependence of the QIF:} we designed a 1.5MHz QIF (250 kHz bandwidth) for varying phases of the filter (The FIR low-pass filter is multiplied by a $\cos(2\pi f_0 t_{\rm sym} + \phi_i)$ where $\phi_i$ is stepped from -$\pi$ to $\pi$. The false-colour plot shows Gain versus signal frequency (x-axis) and filter phase (y-axis), revealing a vertical pass-band at 1.5MHz and a sinusoidal dependence on the filter phase.
\textbf{b. Phase dependence of the QIF:} For a 1.5MHz pre-designed QIF (250 kHz bandwidth) the phase of the filter is scanned (The FIR low-pass filter is multiplied by a $\cos(2\pi f_0 t_{\rm sym} + \phi_i)$ where $\phi_i$ is stepped from 0 to 2$\pi$. The false-colour plot shows contrast versus signal frequency (x-axis) and filter phase (y-axis), revealing a vertical pass-band at 1.5MHz and a sinusoidal dependence on the filter phase.
\textbf{c. A line cut along a fixed frequency:} Contrast at 1.7MHz extracted from panel b. as a function of filter phase. Experimental data (blue circles) follow the predicted $\sin(\phi)^2$ (orange dashed) but are shifted by $\Delta\phi_{\rm signal}\approx\pi/8$, as indicated by the brown fit. This shift was also electronically measured directly and agreed with our results. A maximum contrast occurs when the QIF phase is aligned with the signal phase.
{\textbf{d. Signal Amplitude detection by the QIF:}} A 4$\mu$s band-pass QIF engineered as a 2MHz band-pass is probed with a 2MHz Signal whose amplitude is swept (x-axis). The measures spin contrast $\langle Z \rangle$ (blue circles) grows approximately sinusoidally with the signal strength and is well reproduced by numerical simulation (light-brown dashed line) and explained theoretically (see Supplementary).
\textbf{e. Design of complex pass-bands:}
The QIF protocol generates a time-domain control field that realizes a dual-bandpass frequency response design, centered at $f_0 = 1.5$ MHz and $2.5$ MHz, demonstrating the ability to implement complex spectral filters from arbitrary target profiles.
\textbf{f. Dual band demonstration:} Implementing the design from panel e. yields the measured response (blue circles) with two clear maxima that coincide with the designed profile (orange line), confirming that the QIF faithfully realises the intended two-frequency bandpass. 
}
\label{fig:fig3}
\end{figure*}

In this section, we demonstrate the implementation of the QIF method using the electronic spin of a single nitrogen-vacancy (NV) center in diamond.  This system serves as a controllable and measurable qubit, allowing us to apply the designed control protocol, the input signal, and read out the resulting filtered signal. The experimental setup (Fig.~\ref{fig:fig2}a) includes a laser for qubit initialization and readout, a microwave control field engineered according to the QIF protocol, and a signal input channel, $f_{\rm in}(t) =\cos{(2\pi f t_{\rm sym})}$, with $t_{\rm sym}=t-t_{\rm f}/2$ (see supplementary for details). In this setup, both the signal and the dominant noise act along the $z$-axis, where the noise is primarily of the $1/f$ type. The QIF is specifically designed to suppress this broadband dephasing noise while transmitting the desired signal component.

Figure~\ref{fig:fig2}b shows the first experimental realization of the Quantum Invariant Filter (QIF), implementing a bandpass frequency response centered at a target frequency by applying the designed control field along the $\sigma_x$ axis.
The experimental results closely match the theoretical design, confirming the first experimental verification of this technique. For comparison, a CPMG sequence is tested (Fig.~\ref{fig:fig2}c), operating in the opposite design direction: the frequency response is fixed by selecting a certain number of $\pi$ pulses for a given protocol time $t_{\rm f}$. The resulting frequency response exhibits a clear central frequency accompanied by undesirable and uncontrollable ripples, highlighting its limitations.

The next demonstration highlights another advantage of the QIF approach, its ability to implement frequency-selective filters at continuously tunable frequencies. We design a smoothly varying array of bandpass filters (Fig.~\ref{fig:fig2}d) and experimentally verify that the measured frequency responses closely follow the continuous design (Fig.~\ref{fig:fig2}e). For comparison, we apply a set of CPMG sequences with different numbers of $\pi$ pulses, which yield only discrete filter frequencies and exhibit characteristic sidelobes (ripples) in the frequency domain, Fig.~\ref{fig:fig2}f. In contrast, QIF enables fine-grained control over both the filter center and shape, free from the harmonic constraints inherent in pulsed protocols.

\begin{figure*}[htbp!]
\includegraphics[width=\textwidth]{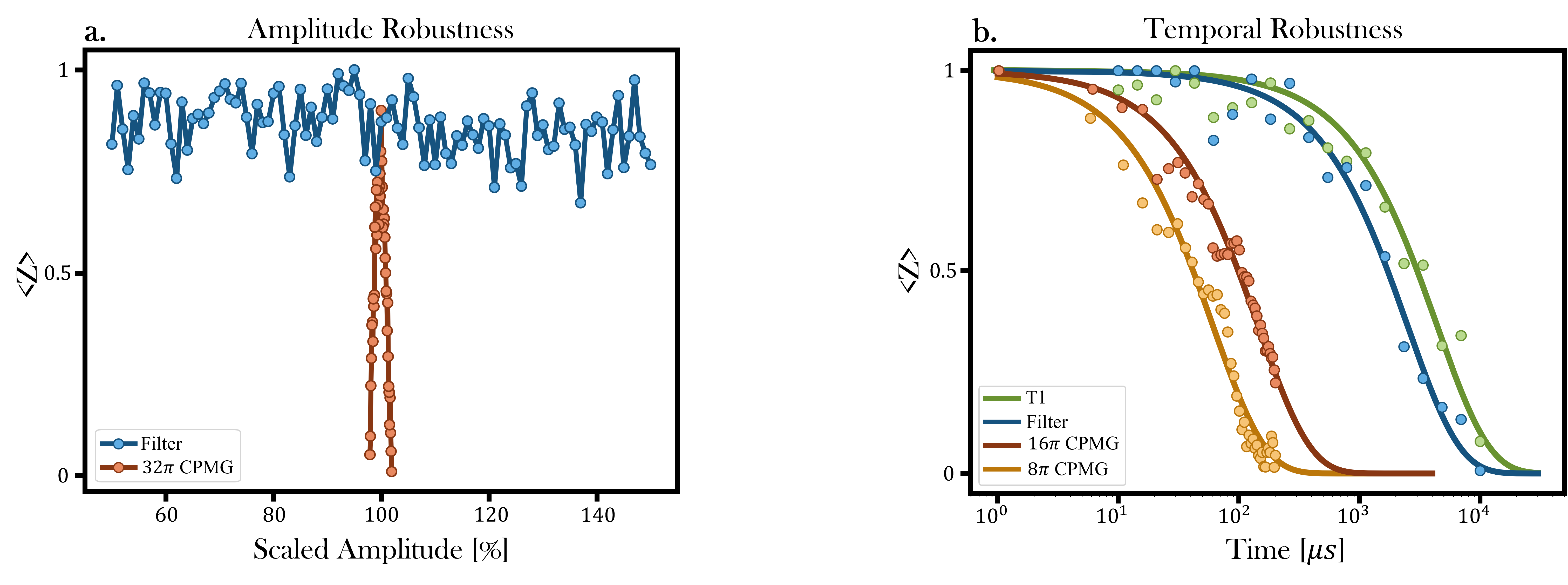}
\caption{\textbf{Robustness of the Quantum-Invariant Filter compared with CPMG:} \textbf{a. Amplitude Robustness:} $\langle Z \rangle$ (y-axis) is plotted as a function of the mw driving amplitude scaled with respect to the calibrated Invariant Field amplitude or the $\pi$-pulse amplitude (CPMG) (x-axis). The blue trace shows the response of a 1 MHz QIF. The orange trace corresponds to a 32-$\pi$ CPMG sequence. The filter maintains above $\gtrsim 80\%$ contrast over a $\pm 50\%$ detuning of the nominal amplitude, whereas the CPMG signal collapses sharply when the pulse departs from the calibrated $\pi$, revealing the superior tolerance of the filter to drive-amplitude errors.
\textbf{b. Temporal Robustness:}
The $\langle Z \rangle$ component(y-axis) versus total evolution time (x-axis, logarithmic scale) for four experiments: 8$\pi$ CPMG (gold, decay time: 0.07 ms), 16$\pi$ CPMG(brown, decay time: 0.14 ms), a 1MHz QIF (blue, decay time: 2.7 ms) and a $T_{1}$ relaxation curve (green, decay time: 4 ms). For each data point the interpulse delay (CPMG) or the filter profile was re-designed for the indicated total time, the $T_1$ data were obtained by initializing the NV in $\ket{0}$, waiting, and reading out $\langle Z \rangle$. The filter outperforms both CPMG variants, extending coherence well beyond 1ms and nearly approaching the intrinsic $T_1$ limit, while the CPMG curves decay almost 2 orders of magnitude faster.}
\label{fig:fig4}
\end{figure*}

The QIF framework permits implementation of lock-in-like filter designs, enabling not only frequency selectivity but also sensitivity to the signal’s phase~\cite{kotler2011single}. Figure~\ref{fig:fig3}a illustrates filter designs with distinct phase responses when driven by a $\sin(2\pi f t_{\rm sym})$ input signal. An evident oscillatory behavior emerges, showing signal null points around phases of $0$, $\pi$, and $-\pi$. Experimentally, as presented in Fig.\ref{fig:fig3}b, each designed filter is realized through the Quantum Invariant Filter (QIF) method, applying the corresponding fields alongside signals of varying frequencies. The oscillatory dependence on phase is clearly observed, although a noticeable phase shift exists between the designed and experimental outcomes. This discrepancy is highlighted more explicitly in Fig.~\ref{fig:fig3}c, where a cross-sectional view at a fixed frequency distinctly reveals the phase offset. We attribute this observed phase shift to the circuit impedance of the signal line, which shifts the signal phase relative to the filter's phase; this conclusion is further corroborated by electronic measurements. This capability to produce a phase-sensitive response effectively provides a method for determining the phase of an unknown signal.

A complementary capability of the filter is its ability to extract signal amplitude information. To investigate signal sensitivity, we scan the filter response as a function of signal amplitude, as illustrated in Fig.~\ref{fig:fig3}d. As previously described, our filter response is second-order in signal amplitude, which is particularly evident at low signal amplitudes where the response exhibits a distinct quadratic, non-linear characteristic. As the signal amplitude increases beyond the perturbative regime, the response transitions to a linear behavior and ultimately saturates. This transition is well-captured by simulation results (brown line), which accurately replicate the overall shape of the experimental response, and is supported by the theory (see Supplementary).

Beyond conventional single-bandpass filters, the QIF method enables the design of complex, multi-band frequency responses, allowing simultaneous probing of multiple spectral features or fingerprints of a system.
To illustrate this flexibility, Fig.~\ref{fig:fig3}e presents a theoretical dual-bandpass frequency filter, designed to simultaneously pass two distinct frequency bands. The experimental validation of this dual-bandpass design, depicted in Fig.~\ref{fig:fig3}f, confirms that the measured frequency response accurately replicates the sophisticated theoretical profile, underscoring the broad design potential and adaptability of the QIF protocol. Indeed, the QIF method provides a generalizable framework for realizing virtually arbitrary frequency-response profiles, greatly exceeding the capabilities of traditional pulse sequences.

Figure~\ref{fig:fig4} evaluates the robustness of the QIF method in the absence of a signal input, focusing on its tolerance to amplitude errors and performance over long evolution times. Fig.~\ref{fig:fig4}a presents the expectation value of the Pauli-$Z$ operator, $\langle Z \rangle$, measured at the end of the protocol as a function of the scaled drive amplitude. The QIF protocol (blue) exhibits remarkable robustness, maintaining $\langle Z \rangle \gtrsim 0.8$ across a broad $\pm 50\% $ amplitude range. In contrast, the CPMG sequence with 32$\pi$ pulses (orange) shows a sharp loss of fidelity away from perfect $\pi$ calibration, highlighting its sensitivity to drive amplitude errors. Fig.~\ref{fig:fig4}b explores the stability of each protocol over time. Here, $\langle Z \rangle$ is plotted versus total evolution time for QIF, 8$\pi$ and 16$\pi$ CPMG sequences, and a $T_1$ relaxation curve. While the QIF Hamiltonian is explicitly time-dependent and not directly comparable to $T_1$-limited free evolution, it nonetheless preserves its desired state over significantly longer durations than CPMG, whose signals decay almost two orders of magnitude faster. These findings establish QIF as a robust and flexible control approach, with enhanced resilience to amplitude fluctuations and environmental noise filtering, compared to traditional pulse-based methods.

\section*{discussion}

Our technique directly translates a designed frequency response into the fields required to achieve this response in the quantum system. This capability significantly expands quantum control capabilities by allowing the precise tailoring of frequency responses to match targeted noise spectra, providing exact control over frequencies, phases, and amplitudes on demand. This approach exhibits superior amplitude robustness compared to traditional methods like the CPMG sequence, making it especially advantageous for manipulating ensembles of NV centers where the amplitude of the control can vary along the sample. Moreover, the QIF method facilitates detailed noise spectroscopy, providing valuable insights into environmental noise characteristics that can inform further refinements in quantum device design. Its versatility ensures broad applicability across various experimental platforms, including nuclear magnetic resonance (NMR), superconducting qubits, and trapped ions. In particular, as was recently demonstrated,  continuous control techniques have been effectively applied to trapped ion~\cite{frey2017application}, and superconducting~\cite{hecht2025beating} systems. Our approach can significantly extend such methods by enabling different platform experiments to directly design filter functions starting from the frequency domain, thus providing enhanced control and performance. Additionally, the QIF method demonstrates remarkable sensitivity and effectiveness in detecting and controlling weak signals, expanding its potential applications in precision quantum sensing and measurement tasks. Future research could explore extending QIF capabilities to multi-qubit systems and real-time adaptive quantum error correction.

\begin{acknowledgments}
This research was supported by the ISF Grants No.~1364/21, 3105/23, 700/22, and 702/22. This research has been generously supported by Dr. Arik Carasso, Honorary Doctor.

\end{acknowledgments}

%

\clearpage
\onecolumngrid  
\section*{Supplementary Information}

\addcontentsline{toc}{section}{Supplementary Information}

\renewcommand{\theequation}{S\arabic{equation}}
\renewcommand{\thefigure}{S\arabic{figure}}
\renewcommand{\thesection}{S\arabic{section}}

\setcounter{section}{0}
\setcounter{equation}{0}
\setcounter{figure}{0}

\section{QIF theory}\label{sec:theorsupp}

Below, we present the theoretical framework underlying the filtering technique proposed in the main text. We summarize the foundational elements of the reverse-engineering approach for constructing the control Hamiltonian, as well as the system's response to external perturbations. We then explain how this response can be systematically tailored to implement both signal and phase filtering.

\subsection{Dynamical Invariants theory}\label{sec:Invtheor}

Let us consider a quantum system driven by classical, deterministic fields that modulate its energy in time, such that its dynamics are governed by a time-dependent Hamiltonian operator $H_{\text{o}}(t)$. The time evolution of a general time-dependent operator $O(t)$ is described by the Heisenberg equation of motion:
\begin{equation}
\frac{\mathrm{d}O(t)}{\mathrm{d} t} = i[H_{\text{o}}(t), O(t)] + \frac{\partial O(t)}{\partial t}.
\end{equation}
Dynamical invariants~\cite{lewis69,Dodonov89,chen2011lewis} are defined as Hermitian, time-dependent operators in the Schrödinger picture, $I(t)$, whose total time derivative vanishes:
\begin{equation}\label{eq:AppHeis}
i [H_{\text{o}}(t), I(t)] + \frac{\partial I(t)}{\partial t} = 0.
\end{equation}

We assume the system has a finite-dimensional Hilbert space, i.e., $\text{dim}(\mathcal{H}_{\text{Hilb}}) = N_{\text{Hilb}}$. Unlike constants of motion, dynamical invariants are explicitly time-dependent; however, Eq.~\eqref{eq:AppHeis} ensures the existence of a complete set of time-dependent eigenvectors $\{\ket{\phi_n(t)}, n = 1, \dots, N_{\text{Hilb}}\}$ with fixed eigenvalues $\lambda_n$, such that
\begin{equation}
I(t) = \sum_{j=1}^{D} \lambda_j \ketbra{\phi_j(t)}.
\end{equation}
The eigenvalues $\lambda_j$ remain constant throughout the evolution. Moreover, the populations of a quantum state $\rho(t)$ in the eigenbasis $\{\ket{\phi_n(t)}\}$ are conserved:
\begin{equation}\label{eq:popres}
\dot{\rho}_{nn}(t) = \frac{\mathrm{d}}{\mathrm{d}t} \bra{\phi_n(t)} \rho(t) \ket{\phi_n(t)} = 0.
\end{equation}
In contrast, the off-diagonal elements (coherences) in this basis evolve as~\cite{levy2018noise}:
\begin{equation}\label{eq:Appcoher}
\frac{\mathrm{d}}{\mathrm{d} t} \rho_{nm}(t) = -i (\dot{\varphi}_n(t) - \dot{\varphi}_m(t)) \rho_{nm}(t),
\end{equation}
where $\varphi_n(t)$ is the Lewis-Riesenfeld (LR) phase, defined as
\begin{equation}\label{eq:AppPhase}
\dot{\varphi}_n(t) = \bra{\phi_n(t)} \left( i \frac{\partial}{\partial t} - H_{\text{o}}(t) \right) \ket{\phi_n(t)}.
\end{equation}
From Eqs.~\eqref{eq:popres},~\eqref{eq:Appcoher}, and~\eqref{eq:AppPhase}, it follows that the evolution operator under $H_{\text{o}}(t)$ can be expressed as
\begin{equation}\label{eq:Invevol}
U_{\text{o}}(t, t_0) = \mathcal{T} e^{-i \int_{t_0}^{t} H_{\text{o}}(t') \mathrm{d}t'} = \sum_{n=1}^{D} e^{i (\varphi_n(t) - \varphi_n(t_0))} \ketbra{\phi_n(t)}{\phi_n(t_0)},
\end{equation}
where $\mathcal{T}$ denotes the time-ordering operator. These features, i.e., population preservation, phase accumulation, and invariant structure, are central to the filtering strategy developed in this work.

Given initial conditions for $I(t)$ and a reference operator basis, Eq.~\eqref{eq:AppHeis} can be used to derive a differential system for $I(t)$ when $H_{\text{o}}(t)$ is specified. Conversely, as discussed in the main text, Eq.~\eqref{eq:AppHeis} can also be employed in reverse to determine the Hamiltonian $H_{\text{o}}(t)$ that generates a desired evolution. This flexibility enables steering the system toward a target state $\rho_{\text{tar}}$ .

This reverse-engineering approach exploits the freedom in choosing $I(t)$: by selecting an appropriate form of the invariant, one can derive families of control Hamiltonians $H_{\text{o}}(t)$ that implement a desired quantum task. A common example involves imposing frictionless boundary conditions:
\begin{equation}\label{eq:fricless}
[H_{\text{o}}(t_a), I(t_a)] = 0,
\end{equation}
where $t_a = t_0, t_{\text{f}}$, and the control protocol spans the interval $[t_0, t_{\text{f}}]$. Enforcing Eq.~\eqref{eq:fricless} ensures that for a closed system the state evolves from an eigenstate of $H_{\text{o}}(t_0)$ to an eigenstate of $H_{\text{o}}(t_{\text{f}})$ with unit fidelity.

As shown below, this framework can be extended further to engineer the system's response to weak external fields, enabling the implementation of signal and phase filtering through reverse-engineered invariant dynamics.


\subsection{SU(2) invariant}
\label{sec:SU2inv}
Eq.~\eqref{eq:AppHeis} is pivotal to define dynamical invariants. On the other hand, it is of limited use in the absence of additional information on the operators $(I(t),H_{\text{o}}(t))$, i.e., on the physical system to be used. In the following, we assume $H_{\text{o}}(t)$ belongs to the $\text{su}(2)$ Lie algebra \cite{d2007introduction,gilmore_2008}. Moreover, we also set $N_{\text{Hilb}}=2$, i.e., the Hilbert space describes a two-level system (TLS).
In this case, the Hamiltonian can be written as a linear combination of generators of the SU2 group, that is, 
\beq\label{eq:HamSU2}
H_{\text{o}}(t)= {\bf h}(t)\cdot {\pmb \sigma}= h_{1}(t)\frac{\sigma_{x}}{2} + h_{2}(t)\frac{\sigma_{y}}{2} + h_{3}(t)\frac{\sigma_{z}}{2}.
\eeq
We also assume that the operator $I(t)$ belongs to the same algebra. It follows that it can be usefully parametrized in terms of an euclidean vector of unit norm $\mathcal{I}(t)$, i.e.,
\beq\label{eq:InvSU2}
I(t)= {\bf \mathcal{I}}(t)\cdot {\pmb \sigma}= \qty(\mathcal{I}_{1}(t)\frac{\sigma_{x}}{2} + \mathcal{I}_{2}(t) \frac{\sigma_{y}}{2} + \mathcal{I}_{3}(t) \frac{\sigma_{z}}{2} ),
\eeq
where at each point in time $\norm{\mathcal{I}(t)}=1$. As a consequence, once they are known at each point in time, the $\mathcal{I}(t)$ components determine the instantaneous basis $\ket{\phi_{j}(t)}$, as well as the LR invariant phase. We choose as a starting reference basis the computational basis $\{\ket{0},\ket{1}\}$. Then, the instantaneous eigenstates $\ket{\phi_{j}(t)}$ can be labeled with the corresponding eigenvalues, i.e., $\lambda_{\pm}=\pm \frac{1}{2} \norm{\mathcal{I}}$  and read
\begin{align}\label{eq:eigenSU2}
\ket{\phi_{-}(t)}&=-\frac{\mathcal{I}_{1}(t) - i\mathcal{I}_{2}(t)}{\norm{\mathcal{I}}\sqrt{2(1+ \mathcal{I}_{3}/\norm{\mathcal{I}})}}\ket{0} + \sqrt{\frac{1}{2}\qty(1 +\frac{\mathcal{I}_{3}}{\norm{\mathcal{I}}})}  \ket{1},\nonumber \\
\ket{\phi_{+}(t)}&=\frac{\mathcal{I}_{1}(t) - i \mathcal{I}_{2}(t)}{\norm{\mathcal{I}}\sqrt{2(1-\mathcal{I}_{3}/\norm{\mathcal{I}})}}\ket{0} + \sqrt{\frac{1}{2}\qty(1 -\frac{\mathcal{I}_{3}}{\norm{\mathcal{I}}})} \ket{1}.
\end{align}

Eqs.~\eqref{eq:eigenSU2} are well-posed provided that $\mathcal{I}_{3}/\norm{\mathcal{I}}\neq \pm 1$. 
Moreover, combining Eq. \eqref{eq:AppPhase} with Eq.\eqref{eq:HamSU2} and Eq. \eqref{eq:eigenSU2},
we also find that 
\beq\label{eq:LRderi}
\dot{\varphi}_{\pm}(t)= \frac{1}{2(1\mp \mathcal{I}_{3}(t))}\qty(\mathcal{I}_{1}(t) \dot{\mathcal{I}}_{2}(t)-\mathcal{I}_{2}(t)\dot{\mathcal{I}}_{1}(t))\mp \frac{1}{2} {\bf h}(t)\cdot \mathcal{I}(t).  
\eeq

The structure of the SU2 group also sets the link between the invariant components and the Hamiltonian control fields $h_{i}(t)$. Indeed, plugging Eqs.~\eqref{eq:HamSU2}, ~\eqref{eq:InvSU2} into Eq.~\eqref{eq:AppHeis}, a system of differential equation can be derived that reads
\begin{align}\label{eq:system}
\dot{\mathcal{I}}(t)=\mathcal{M}(t)\cdot \mathcal{I}(t) \mbox{, }
\mathcal{M}(t)=
\begin{pmatrix}
0 & h_{3}(t) & -h_{2}(t)\\
-h_{3}(t) & 0 & h_{1}(t)\\
h_{2}(t) & -h_{1}(t) & 0
\end{pmatrix}
\end{align}
It is important to notice that, when interpreted as an algebraic system, Eq.~\eqref{eq:system} cannot be inverted. However, from the property of the conservation of the norm of $\mathcal{I}$, i.e., $\sum_{i=1}^{D}\dot{\mathcal{I}}_{i} \mathcal{I}_{i}=0$, we can always find at least an index $k^{*}$ such that the fields $h_{i}(t)$ in terms of the $\mathcal{I}$ can be expressed as follows
\beq\label{eq:partinv}
h_{i}(t)=-\sum_{j=1}^{D} \varepsilon_{ij k^{*}}\frac{\dot{\mathcal{I}}_{j}}{\mathcal{I}_{k^{*}}} + h_{k^{*}}\frac{\mathcal{I}_{i}}{\mathcal{I}_{k^{*}}}
\eeq
where we denoted with $\varepsilon_{ijk}$ the Levi-Civita tensor. It follows that, when a suitable set of boundary conditions are chosen, we can express two control fields in terms of all the invariant components $\mathcal{I}$, thus allowing for a reverse engineering strategy of the Hamiltonian $H_{\text{o}}(t)$.

\subsubsection{Parametrization of the invariant and boundary conditions}

To set the stage of our reverse-engineered protocols, we set $h_{2}(t)=0$ at each point in time. Then, following Eq.~\eqref{eq:partinv}, we can set $k^{*}=3$ and, using the conservation of the norm of $\mathcal{I}$, we can write
\beq\label{eq:partinv}
h_{1}(t)=\frac{\dot{\mathcal{I}}_{3}(t)}{\mathcal{I}_{2}(t)}\mbox{, }  h_{3}(t)=-\frac{\dot{\mathcal{I}}_{1}(t)}{\mathcal{I}_{2}(t)}. 
\eeq
Moreover, it is convenient to adopt a parametrization of $\mathcal{I}(t)$ in terms of two independent auxiliary functions $(\alpha(t),\beta(t))$ which can be interpreted as angles on the SU2 sphere. We thus rewrite Eq.~\eqref{eq:InvSU2} as follows
\beq\label{eq:InvSU2def}
I(t)= \qty(-\cos \alpha(t)\frac{\sigma_{x}}{2} + \sin \alpha(t) \sin \beta(t) \frac{\sigma_{y}}{2} +\sin \alpha(t) \cos \beta(t) \frac{\sigma_{z}}{2}),
\eeq
Notice that, compared to the usual convention on the azimuthal and polar angles, we parametrize $\mathcal{I}(t)$ in terms of a $\pi/2$-rotated vector around the $\hat{y}$ axis. From Eqs.\eqref{eq:partinv} \eqref{eq:InvSU2def}, setting $h_{1}(t)=-\varepsilon(t)\mbox{, } h_{3}(t)=\Delta(t)$ we find
\beq\label{eq:fieldeq}
\varepsilon(t)=\dot{\beta}(t) -\frac{\dot{\alpha}(t)}{\tan \alpha(t) \tan \beta(t) }, \qquad \Delta(t)=-\frac{\dot{\alpha}(t)}{\sin \beta(t)},
\eeq
as reported in the main text. Following Eq.~\eqref{eq:fieldeq}, an arbitrary choice of the auxiliary functions $(\alpha(t),\beta(t))$ allows to compute the Hamiltonian control fields linked to the evolution operator in Eq.~\eqref{eq:Invevol}. As a consequence, if the system is prepared at the initial time $t=t_{0}$ in a pure state $\ket{\psi(t_{0})}=\ket{\phi_{i}(t_{0})}$ then, up to a global phase, it will be transported in the same time-dependent eigenstate during the dynamics. On the other hand, for initial pure states that are linear superpositions of the eigenvectors $\ket{\phi_{i}(t_{0})}$, the system state will go through linear superpositions of the instantaneous eigenvectors, conserving the initial populations and developing relative phases ruled by Equations \eqref{eq:Appcoher},~\eqref{eq:AppPhase},~\eqref{eq:Invevol}.   

The last step required is to set the properties of the fields at initial and final times of the protocol. In turn, this is also needed to set the target state of the reverse-engineered control. We thus impose the frictionless conditions in Eq. ~\eqref{eq:fricless} at initial and final time. It is easy to find that Eq.~\eqref{eq:fricless} consists of 6 equations involving the invariant and the control fields. Following the parametrization in Eq.~\eqref{eq:InvSU2def}, provided that the control fields are finite, a suitable choice to fulfill Eq.~\eqref{eq:fricless} is the following
\beq\label{eq:bounddef}
\alpha(0)=\alpha(t_{\text{f}})=\pi,  \qquad \dot{\alpha}(0)=\dot{\alpha}(t_{\text{f}})=0, \qquad \beta(0)=\beta(t_{\text{f}})=-\frac{\pi}{2}. 
\eeq
Notice also that Eq.~\eqref{eq:bounddef} implies $\Delta(t_{a})=0$, i.e., $(H_{\text o}(t),I(t))$ are proportional to $\sigma_{x}$ at the start and at the end of the protocol. Furthermore, the boundary conditions set for $\beta(t)$ ensure that Eqs.~\eqref{eq:fieldeq} are well-posed for $t=t_{0},t_{\text f}$. Except for the endpoint times, for $t_{\text o}< t< t_{\text f}$ the auxiliary fields $(\alpha(t),\beta(t))$ can thus be chosen to take arbitrary shapes, provided that Eqs.~\eqref{eq:fieldeq} are well-posed.


\subsection{From Lab to Rotating Frame}
We start from the Hamiltonian in the lab frame, which models the interaction of the microwave fields with the NV center. The Hamiltonian of the driven NV center is written as follows
\beq\label{eq:Labframe}
H_{\text{lab}}(t)=\frac{\omega_{\text{NV}}}{2}\sigma_z + \varepsilon(t)\sin (\omega_d t +\phi_d)\sigma_{x}, 
\eeq
where $\omega_{\text{NV}} \simeq 1.5 \text{GHz}$ is the energy gap of the two levels of the NV center and $\omega_{d}$ is the frequency of the microwave field and $\varepsilon(t)$ is the controlled field amplitude, which is slowly changing with time as compared with $T_d=2\pi/\omega_d$.
We aim to control the dynamics of the NV center in the rotating frame with frequency $\omega_{\text{NV}}$ around the $z$ axis. Indeed, applying the conventional transformation $U_r=e^{iH_{\text{NV}}t}$, the Hamiltonian in the rotating wave takes the form 
\beq\label{eq:rotframe}
H_{r}=i\frac{\mathrm{d}U_{r}}{\mathrm{d}t} U^{\dagger}_{r} + U_{r}H_{\text{lab}}U^{\dagger}_{r}=\varepsilon(t)\sin (\omega_d t +\phi_d)\qty[\cos (\omega_{\text{NV}}t) \sigma_x -\sin( \omega_{\text{NV}}t) \sigma_y].
\eeq
From Eq.\eqref{eq:rotframe} it is easy to see that, neglecting the fast-rotating terms with frequency $\omega_d +\omega_{\text{NV}}$, choosing the driving field to be in resonance with the gap and setting $\phi_d=\pi/2$, we obtain
\beq\label{eq:rotframe}
H_{r}(t)=\frac{\varepsilon(t)}{2}\sigma_{x}.
\eeq
\subsection{Response to external fields }\label{sec:response}

We consider the reverse-engineered control fields as detailed in Sec.~\ref{sec:SU2inv}, i.e., following a given choice of the auxiliary fields $(\alpha(t),\beta(t))$ the control Hamiltonian $H_{\text{o}}(t)$ is known. Below, we focus on the response of the reverse-engineered system to an arbitrary time-dependent external field acting on top of the control fields (these could be signals we wish to detector or a noise term). 
The TLS Hamiltonian thus reads 
\beq\label{eq:pert1}
H(t)=H_{\text{o}}(t) + V(t), 
\eeq
where $V(t)=\delta f_{\text{in}}(t) V$, $\delta $ is a sufficiently small amplitude as compared with the control fields, $f_{\text{in}}(t)$ is an arbitrary function of time and $V$ can be written as a linear combination of SU(2) operators. For our purpose, it suffices to work with pure states. In order to study the response, we take advantage of the analytically-closed form of the solution in the limit $\delta =0$, employing standard perturbation theory.  Although the perturbative approach is expected to work only in limited regions of the system parameters, it offers useful insights to understand the basic mechanism underlying the QIF filter, and it can be validated by means of numerical simulations (see Sec.~\ref{subsec:accuracy}).
We adopt the interaction picture w.r.t. the Hamiltonian $H_{\text{o}}(t)$, where we consider Eq.~\eqref{eq:Invevol} as the unperturbed evolution operator. 
Then, by denoting with $U(t,t_{0})=\mathcal{T}e^{-i\int_{t_{0}}^{t}H(s) \mathrm{d}s}$ the evolution operator corresponding to the whole Hamiltonian in the Schr\"odinger picture, the evolution operator in the interaction picture reads $U_{\text{I}}(t,t_{0})=U^{\dagger}_{\text{o}}(t,t_{0})U(t,t_{0})$. It is a solution of the following differential equation
\beq\label{eq:pert2}
i \frac{\partial}{\partial  t}U_{\text{I}}(t,t_{0})=V_{\text{I}}(t)U_{\text{I}}(t,t_{0}),
\eeq
with $V_{\text{I}}(t)=U^{\dagger}_{\text{o}}(t,t_{0})V(t)U_{\text{o}}(t,t_{0})$ and obeys the initial condition $U_{\text{I}}(t_{0},t_{0})=\mathbb{1}$. The solution of Eq.~\eqref{eq:pert2} is conventionally written in terms of the Dyson series. In order to describe a suitable approximation of our filter, it will be sufficient to truncate the Dyson series including the terms up to the second order. Thus, we can write 
\beq\label{eq:pert3}
U_{\text{I}}(t,t_{0})= \mathbb{1} -i \int_{t_{0}}^{t} V_{\text{I}}(s)U_{\text{I}}(s,t_{0})\mathrm{d}s\simeq \mathbb{1} -i \int_{t_{0}}^{t} V_{\text{I}}(s)\mathrm{d}s -\int_{t_{0}}^{t}\mathrm{d}s\int_{t_{0}}^{s}\mathrm{d}s^{\prime}V_{\text{I}}(s)V_{\text{I}}(s^{\prime}),
\eeq
where, due to time ordering operator, $t_{0}<s^{\prime}<s <t$. We compute the correction to the ket state in the Schr\"odinger picture arising from the action of the perturbation as follows
\beq\label{eq:pert4}
\ket{\psi(t)}=U_{\text{o}}(t,t_{0}) U_{\text{I}}(t,t_{0})\ket{\psi(t_{0})} \simeq U_{\text{o}}(t,t_{0})\qty(\mathbb{1} -i\int_{t_{0}}^{t}V_{\text{I}}(s)\mathrm{d}s - \int_{t_{0}}^{t}\mathrm{d}s\int_{t_{0}}^{s}\mathrm{d}s^{\prime}V_{\text{I}}(s)V_{\text{I}}(s^{\prime}))\ket{\psi(t_{0})}.
\eeq
It is worth stressing that, in the absence of a closed-form analytical solution for $U_{\text{o}}(t,t_{0})$, Eq.~\eqref{eq:pert4} would be of limited use. On the contrary, from the properties of the dynamical invariant, i.e., Eqs.~\eqref{eq:AppPhase}, \eqref{eq:Invevol},\eqref{eq:InvSU2def}, it is easy to compute Eq.~\eqref{eq:pert4} and, at least in principle, any higher-order perturbation term. Indeed, from the perspective of reverse-engineering of the filter fields, as detailed in Sec.~\ref{sec:FIR}, the invariant parameters $(\alpha(t),\beta(t))$ are designed to achieve the desired frequency response.
\subsubsection{First-order response}
By plugging Eq.~\eqref{eq:Invevol} into Eq.~\eqref{eq:pert4}, after some algebra we get the first-order correction to the state that reads
\begin{multline}\label{eq:pert7}
\ket{\psi(t)}\simeq\ket{\psi_{\text{o}}(t)} + \ket*{\chi^{(1)}(t)}\mbox{, } \ket*{\chi^{(1)}(t)}= -i\Big[\Big(\bra{\phi_{+}(t_{0})}\ket{\psi(t_{0})}L_{11}(t) + \bra{\phi_{-}(t_{0})}\ket{\psi(t_{0})} L_{12}(t)\Big)\cdot\\\cdot e^{i(\varphi_{+}(t)-\varphi_{+}(t_{0}))}\ket{\phi_{+}(t)}+
\Big(\bra{\phi_{+}(t_{0})}\ket{\psi(t_{0})} L_{21}(t) +\bra{\phi_{-}(t_{0})}\ket{\psi(t_{0})} L_{22}(t)\Big) e^{i(\varphi_{-}(t)-\varphi_{-}(t_{0})))}\ket{\phi_{-}(t)}\Big],
\end{multline}
with 
\begin{align}\label{eq:pert8}
L_{11}(t)&=\int_{t_{0}}^{t}\mathrm{d}s\bra{\phi_{+}(s)}V(s)\ket{\phi_{+}(s)},L_{22}(t)=\int_{t_{0}}^{t}\mathrm{d}s\bra{\phi_{-}(s)}V(s)\ket{\phi_{-}(s)},\nonumber\\
L_{12}(t)&=\int_{t_{0}}^{t}\mathrm{d}s\bra{\phi_{+}(s)}V(s)\ket{\phi_{-}(s)}e^{i(\varphi_{-}(s)-\varphi_{+}(s)-(\varphi_{-}(t_{0})-\varphi_{+}(t_{0})))},\nonumber\\
L_{21}(t)&=\int_{t_{0}}^{t}\mathrm{d}s\bra{\phi_{-}(s)}V(s)\ket{\phi_{+}(s)}e^{i(\varphi_{+}(s)-\varphi_{-}(s)-(\varphi_{+}(t_{0})-\varphi_{-}(t_{0})))}.
\end{align}
It should be noticed that $\ket{\psi_{\text{o}}(t)}$ denotes the ket state as ruled by $H_{\text{o}}(t)$ alone. Furthermore, for any given choice of the invariant auxiliary fields, the correction to the reverse-engineered unitary dynamics depends on the operator $V(t)$, as well as on the initial conditions.  
To derive the results reported in the main text, we assume the external field to act as a modulation of the bias field along $\hat{z}$, i.e., we take $V(t)=\delta f_{\text{in}}(t) \sigma_{z}/2$. Although the derivation holds for arbitrary fields $f_{in}(t)$, to elucidate the working principle of the FIR we will be dealing with periodic functions of time. Moreover, we make use of the parametrization reported in Eqs.~\eqref{eq:InvSU2def},~\eqref{eq:bounddef}, and we set $t_{0}=0$.   

It is useful to notice that 
\begin{align}\label{eq:pert10}
 \bra{\phi_{+}(s)}V(s)\ket{\phi_{+}(s)}&=-\bra{\phi_{-}(s)}V(s)\ket{\phi_{-}(s)}=\frac{\delta}{2}f_{\text{in}}(s)\sin \alpha(s)\cos \beta(s),\nonumber\\
 \bra{\phi_{+}(s)}V(s)\ket{\phi_{-}(s)}&=-\frac{\delta }{2}f_{\text{in}}(s)\sqrt{1-\sin^2 \alpha(s) \cos^2 \beta(s)}.
\end{align}
Plugging Eqs.~\eqref{eq:pert8},~\eqref{eq:pert10}, into Eq.~\eqref{eq:pert7}, the expectation values of the arbitrary Pauli operator $\sigma_{i}$ can be computed by keeping only the first-order contributions in $\delta $ as follows
\beq\label{eq:pert11}
\bra{\psi(t)}\sigma_{i}\ket{\psi(t)}\simeq  \bra{\psi_{\text{o}}(t)}\sigma_{i}\ket{\psi_{\text{o}}(t)}+\bra*{\chi^{(1)}(t)}\sigma_{i}\ket{\psi_{\text{o}}(t)}+\bra{\psi_{\text{o}}(t)}\sigma_{i}\ket*{\chi^{(1)}(t)} +o(\delta^2) .
\eeq
For $t=t_{\text{f}}$  we thus get
\begin{align}\label{eq:pert12}
\ev{\sigma_{x}(t_{\text{f}})}&\simeq  
\ev{\sigma_{x}(t_{\text{f}})}_{\text{o}} -\delta \int_{0}^{t_{\text{f}}}\mathrm{d}s f_{\text{in}}(s) \sqrt{1-\sin^2 \alpha(s)\cos^2 \beta(s)}\sin(\Delta\varphi(s) -\Delta\varphi(t_{0})),\nonumber\\
\ev{\sigma_{y}(t_{\text{f}})}&\simeq  \ev{\sigma_{y}(t_{\text{f}})}_{\text{o}} + \delta \mathcal{A}_{1}(t_{\text{f}})\cos(\Delta \varphi(t_{\text{f}})-\Delta \varphi(t_0)),\nonumber\\
\ev{\sigma_{z}(t_{\text{f}})}&\simeq  \ev{\sigma_{z}(t_{\text{f}})}_{\text{o}} +\delta \mathcal{A}_{1}(t_{\text{f}})\sin(\Delta\varphi(t_{\text{f}})-\Delta\varphi(t_{0})),
\end{align}
where $\Delta\varphi(t)=\varphi_{+}(t)-\varphi_{-}(t)$ is the difference in the LR invariant phase at time $t$, and the quantity $\mathcal{A}_{1}(t_{\text{f}})$, which describes the first-order response, reads
\beq\label{eq:ampliagain}
\mathcal{A}_{1}(t_{\text{f}})=\int_{t_{0}}^{t_{\text{f}}}\mathrm{d}s f_{\text{in}}(s)\sin \alpha(s) \cos \beta(s).
\eeq
Moreover, from the solution of the unperturbed reverse-engineered dynamics it follows 
\beq\label{eq:pert13}
\ev{\sigma_{x}(t_{\text{f}})}_{\text{o}}=\ev{\sigma_{y}(t_{\text{f}})}_{\text{o}}=0,\ev{\sigma_{z}(t_{\text{f}})}_{\text{o}}=\cos(\Delta\varphi(t_{\text f})-\Delta\varphi(t_{0})).
\eeq 

It is thus easy to see that the first-order response to the signal field can be reverse-engineered by tailoring the time-dependence of the invariant fields $(\alpha(t),\beta(t))$ and, in turn, the LR dynamical phase $\varphi_{\pm}(t)$. Notice that, due to the special form of Eq.~\eqref{eq:pert13}, desired features of the linear response could also be achieved by tuning the endpoint-values of the dynamical phase. 
\subsubsection{Second-order response}
The design of the first-order response may not be sufficient to achieve significant filtering effects, so second-order terms in Eq.~\eqref{eq:pert11} are considered as well.
We thus take into account second-order terms, as written in Eq.~\eqref{eq:pert4}. Again, inserting the analytical expression in Eq.~\eqref{eq:Invevol} into Eq.~\eqref{eq:pert4}, it can be found that the additional contribution has the form 
\beq\label{eq:ketsecond}
\ket*{\chi^{(2)}(t)}=\sum_{j=\pm}f_{j}(t)e^{i(\varphi_{j}(t)-\varphi_{j}(t_{0}))}\ket{\phi_{j}(t)},
\eeq
where
\begin{multline}\label{eq:coeffsec}
f_{j}(t)=-\sum_{kl}\int_{t_{0}}^{t}\mathrm{d}s\int_{t_{0}}^{s}\mathrm{d}s^{\prime}e^{i(\varphi_{k}(s)-\varphi_{j}(s)) -i(\varphi_{k}(t_0)-\varphi_{j}(t_0))}e^{i(\varphi_{l}(s^{\prime})-\varphi_{k}(s^{\prime})) -i(\varphi_{l}(t_0)-\varphi_{k}(t_0))}\cdot\\\cdot \bra{\phi_{j}(s)}V(s)\ket{\phi_{k}(s)} \bra{\phi_{k}(s^{\prime})}V(s^{\prime})\ket{\phi_{l}(s^{\prime})}\braket{\phi_{l}(t_0)}{\psi(t_{0})}.
\end{multline}
Notice that Eq.~\eqref{eq:ketsecond},~\eqref{eq:coeffsec} involve the sum over different terms, which require integrating products of functions in Eq. ~\eqref{eq:pert10} and dynamical phase differences. Indeed, the perturbing field $V(t)$ displays diagonal and off-diagonal nonzero elements in the basis $\ket{\phi_{j}(t)}$, both contained in the integrals required. The combination of these different terms may not be easy to handle, and the overall response is expected to be determined by first-order effects. However, in Sec.~\ref{sec:FIR}, we will show how the auxiliary control fields $(\alpha(t),\beta(t))$ can be suitably designed to achieved a desired second-order response to external fields and realize a desired filter function.

\subsection{Quantum response and filtering with dynamical invariants}
\label{sec:FIR}

To implement frequency-selective quantum control using the invariant-based reverse engineering framework described in Sec.~\ref{sec:Invtheor}, we design control protocols that evolve over a finite time interval $t_{\text{f}}$. These protocols are intended to tailor the response of the system, specifically, an NV center spin qubit, to external fields in a manner that realizes a desired filter function, as further discussed in Sec.~\ref{sec:response}. Our approach begins with a target filtering function $\mathcal{H}(\omega)$ in the frequency domain and aims to encode this response into the time-domain dynamics of the system using reverse-engineered dynamical invariants.

To achieve this, we construct time-localized filter functions that vanish at both endpoints of the protocol, consistent with the frictionless boundary conditions required by the invariant method. Finite impulse response (FIR) filters provide a natural choice for this purpose: they are intrinsically finite in duration and can be designed to exhibit well-controlled spectral characteristics. In particular, we focus on the design of band-pass FIR filters characterized by a center frequency $f_0$ and a cutoff frequency $\nu_c$, which defines the bandwidth. FIR filters are typically implemented in discrete time and defined by a sequence $\mathcal{H}[n]$, where $n = 0, \dots, N_{\text{F}}-1$, and $N_{\text{F}}$ is the number of filter coefficients (or taps). The effective time duration of the filter is $T = (N_{\text{F}} - 1)/F_s$, where $F_s$ is the sampling frequency. This duration is matched to the total protocol time $t_{\text{f}}$ to ensure compatibility with the invariant evolution.

To integrate FIR filtering into the continuous-time dynamics of quantum control, we approximate the discrete impulse response $\mathcal{H}[n]$ by a smooth, continuous function $\mathcal{H}(t)$. This step is essential to ensure that time derivatives appearing in the invariant formalism remain well-defined and free of discontinuities. 

We encode this filter into the invariant framework by choosing the auxiliary functions as
\begin{align}\label{eq:auxFIR}
\alpha(t) &= \pi,\nonumber \\
\beta(t) &= -\frac{\pi}{2} + \arcsin \mathcal{H}(t).
\end{align}
These choices satisfy the required frictionless boundary conditions (Eqs.~\eqref{eq:bounddef}) as long as $\mathcal{H}(0) = \mathcal{H}(t_{\text{f}}) = 0$. As discussed in Sec.~\ref{sec:response}, these auxiliary fields fully determine the system's response to external signals.

An important consequence of this construction is the absence of control fields along the $\hat{z}$-axis.
As a result, the dynamical invariant remains aligned with the $\hat{x}$-direction throughout the evolution and becomes time-independent. The corresponding eigenstates are constant, $\ket{\phi_{\pm}(t)} = \ket{x, \pm}$, which greatly simplifies experimental implementation (see Sec.~\ref{sec:setup}).

The NV center is initialized in the state $\ket{\psi(0)} = \ket{0} = (1/\sqrt{2})(\ket{\phi_+} - \ket{\phi_-})$. Under the action of the engineered Hamiltonian $H_0(t)$, the populations of the invariant eigenstates remain fixed, while their relative phases evolve according to the Lewis-Riesenfeld (LR) phase. Using Eq.~\eqref{eq:AppPhase}, the LR phase for this protocol is
\begin{equation}\label{eq:FIRLR}
\dot{\varphi}_{\pm}(t) = \bra{\phi_{\pm}} i \partial_t - H_0(t) \ket{\phi_{\pm}} = \pm \frac{\dot{\beta}(t)}{2}, \quad \Rightarrow \quad \varphi_+(t) - \varphi_-(t) = \beta(t).
\end{equation}
At the final time $t = t_{\text{f}}$, the relative phase returns to its initial value, and the quantum state is restored to its original form. At intermediate times, however, the system undergoes coherent phase evolution governed by the structure of $\mathcal{H}(t)$, which encodes the designed filter profile. This evolution enables selective sensitivity to external signals, as described in Sec.~\ref{sec:response}, thereby implementing the desired Quantum Invariant Filter. We note that in practice the axillary field $\beta(t)$ can be simplified by setting it to $\beta(t)=-\frac{\pi}{2}+\mathcal{H}(t)$ as discussed in Sec.~\ref{sec:I_filter_implement}


\subsubsection{Second-order response under FIR-modulated invariants}\label{subsec:SecorderFIR}

In order to design the QIF, we restrict to the case of auxiliary fields described in Eqs.~\eqref{eq:auxFIR}. We stress that, although in this limit the invariant in Eq.~\eqref{eq:InvSU2def} reduces to a simple constant of the motion, this does not reduce the generality of our approach, and many more solutions using our method can be found. On the other hand, the advantage of this choice is that, at first order in the external field amplitude $\delta $, the expectation values of $\ev{\sigma_{y}(t)},\ev{\sigma_{z}(t)}$ remain unchanged, i.e., in Eq.~(\ref{eq:ampliagain}) we have $\mathcal{A}_{1}(t_{\text f})=0$. As a result, projective measurement of the two-level system populations at time $t=t_{\text f}$ allows us to probe second-order perturbation effects on the invariant dynamics. 
Starting from Eq.~\eqref{eq:pert3}, we can include second-order effects in the unitary $U_{\text{I}}(t,t_{0})$ due to the signal field, as written in Eqs.~\eqref{eq:ketsecond} and ~\eqref{eq:coeffsec}. After some algebra, under the previous simplifying assumption, we find
\beq\label{eq:pert14}
\ev{\sigma_{z}(t_{\text f})}\simeq \ev{\sigma_{z}(t_{\text f})}_{0} -\frac{\delta^2}{2}\qty[\int_{0}^{t_{\text f}}\mathrm{d}s f_{\text{in}}(s) \sin(\Delta \varphi(s )-\Delta \varphi(0 ))]^2. 
\eeq 
From Eqs.~\eqref{eq:bounddef} and \eqref{eq:FIRLR}, it also follows that $\Delta \varphi(0)=-\pi/2$, so that Eq.~\eqref{eq:pert14} can be easily rewritten as
\beq\label{eq:pert16}
\ev{\sigma_{z}(t_{\text f})}\simeq \ev{\sigma_{z}(t_{\text f})}_{0} -\frac{\delta^2}{2}\qty[\int_{0}^{t_{\text f}}\mathrm{d}s f_{\text{in}}(s) \cos(\Delta \varphi(s ))]^2. 
\eeq 
Note that for  arbitrary input function $f_{\text{in}}(s)$, and under the previously stated assumptions on the auxiliary fields, the symmetry of the second-order response enables a compact expression for the deviation in the expectation value of $\sigma_z$ at the final time:
\begin{equation}\label{eq:pert20}
\langle \sigma_z(t_{\text{f}}) \rangle - \langle \sigma_z(t_{\text{f}}) \rangle_{\text{0}} = -\frac{\delta^2}{2} \mathcal{A}_2^2(t_{\text{f}}) = -\frac{\delta^2}{2} \left[ \int_0^{t_{\text{f}}} \mathrm{d}s f_{\text{in}}(s) \mathcal{H}(s) \right]^2,
\end{equation}
where $\mathcal{H}(t) = \cos (\Delta \varphi(t))$ acts as an effective impulse response function, which depends on the LR phase of the unperturbed system,  $\Delta \varphi(t)=\beta(t)$, as defined in Eq.~\eqref{eq:auxFIR}. This expression links the system’s response to deviations from the ideal final population, i.e., a breakdown of the dynamical invariant governed by Eq.\eqref{eq:AppHeis}, caused by the external perturbation. Crucially, Eq.~\eqref{eq:pert20} enables the design of desired filter functions in the frequency domain, which can then be implemented through corresponding time-domain control fields.

\subsubsection{FIR and convolution structure of the QIF response}\label{subsec:convolution}
The second-order correction to the expectation value $\langle \sigma_z(t_{\rm f}) \rangle$ derived in Eq.~\eqref{eq:pert20} reveals a clear convolution structure. Specifically, the deviation from the unperturbed outcome depends quadratically on a filtered signal amplitude,
\begin{equation}\label{eq_A2}
\mathcal{A}_2(t_{\rm f}) = \int_0^{t_{\rm f}} ds f_{\text{in}}(s)\mathcal{H}(s)
\end{equation}
where $\mathcal{A} \equiv \mathcal{A}_2$ in the main text.
Assuming that $\mathcal{H}(t)$ vanishes outside the control window $[0, t_{\rm f}]$, a requirement consistent with the boundary conditions of the dynamical invariant, and that $\mathcal{H}(s)$ is symmetric about $t_{\rm f}/2$, i.e., $\mathcal{H}(s) = \mathcal{H}(t_{\rm f} - s)$ (which can be enforced by construction), the integral takes the form of a convolution:
\begin{equation}\label{eq:A}
\mathcal{A}(t_{\rm f}) = \left(f_{\text{in}} \ast \mathcal{H} \right)(t_{\rm f}) = \int_{-\infty}^{\infty} ds f_{\text{in}}(s)\mathcal{H}(t_{\rm f} - s).
\end{equation}

This establishes that the QIF protocol acts as a linear filter on the signal, with $\mathcal{H}(t)$ as the temporal filter kernel. In the frequency domain, the filtering process corresponds to a spectral multiplication:
\begin{equation}
\mathcal{A}(\omega) = f_{\text{in}}(\omega)\mathcal{H}(\omega) ,
\end{equation} 
Hence, the output response at the end of the protocol reflects the selective transmission of frequency components defined by the filter shape $\mathcal{H}(t)$, which is determined by the choice of auxiliary field $\beta(t)$ through the invariant dynamics.

\subsubsection{Accuracy of the perturbation approach}\label{subsec:accuracy}

\begin{figure}[h!]
\includegraphics[scale=0.40]{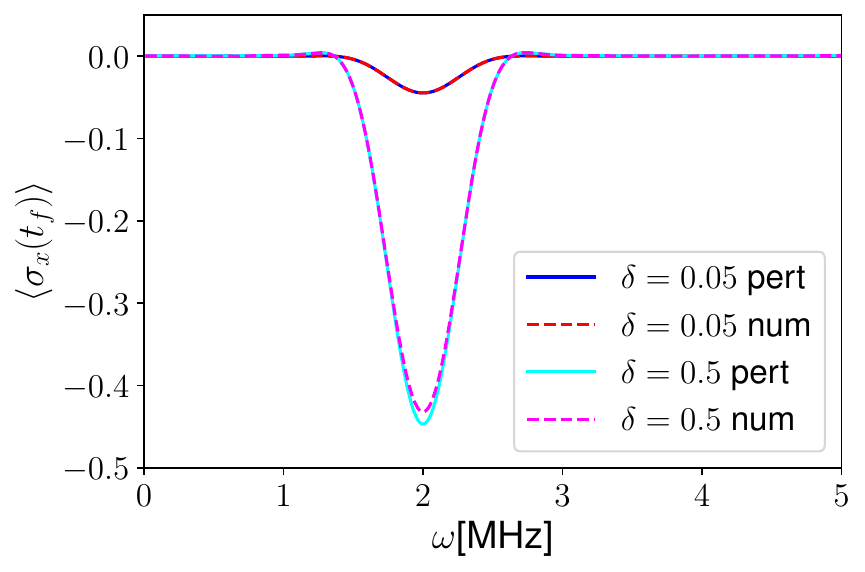}
\includegraphics[scale=0.40]{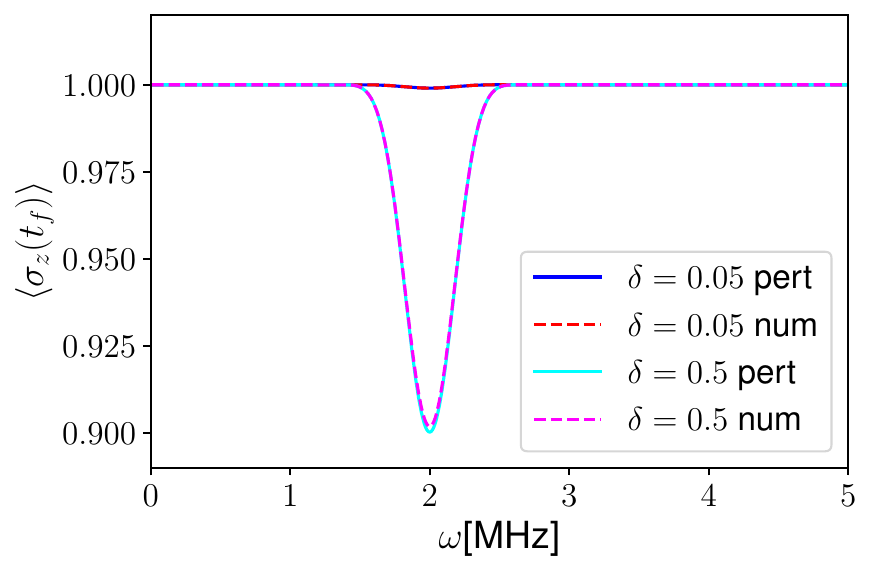}
\caption{Comparison of the QIF response (solid line) with numerical simulations (dashed line) for monochromatic external signal $f_{in}(t)=\cos (2\pi\omega(t-t_{\text f}/2))$, computed against $\omega$ for two different values of the amplitude $\delta =0.05,0.5$, $t_{\text f}=4 \mu s$.}
\label{fig:numvspert}
\end{figure}

The time-dependent perturbation approach is a straightforward tool to model the basic mechanism of the QIF. However, due to the truncation introduced in Eq.~\eqref{eq:pert4}, it is not expected to recover the actual response to the external fields. This holds especially true for increasing values of $\delta \cdot t_{\text f}$. As shown in Fig.~\ref{fig:numvspert}, the comparison with numerical simulations of the dynamics as computed from Eq.~\eqref{eq:pert1} shows the deviations from the exact behavior mainly take place in the peak values of the response, while the desired frequency profile is correctly reproduced. This also holds independently on the FIR filter parameters chosen, i.e., $f_0$. It follows that Eq.~\eqref{eq:pert20} provides a satisfactory approximation of QIF response and can be employed to showcase the QIF capabilities.    

\subsubsection{Magnus expansion}\label{subsec:Magnus}
The results presented in \ref{subsec:SecorderFIR} can be equivalently derived by means of Magnus expansion \cite{blanes2009magnus, kuprov2023optimal, d2024open}. Indeed, the solution of Eq.~\eqref{eq:pert2} can be written as follows
\begin{equation}\label{eq:magnu1}
U_{\text{I}}(t,t_{0})=\mathcal{T}e^{-i\int_{t_{0}}^{t}V_{\text{I}}(t^{\prime}) \mathrm{d}t^{\prime}}=e^{\Omega(t)},
\end{equation} 
where, following Magnus theorem, the exponent can be expressed in terms of an infinite series of operators that reads
\begin{equation}\label{eq:magnu2}
\Omega(t)= -i\int_{t_{0}}^{t}V_{\text{I}}(s)\mathrm{d}s -\frac{1}{2} \int_{t_{0}}^{t}\mathrm{d}s\int_{t_{0}}^{s}\mathrm{d}s^{\prime} \comm{V_{\text{I}}(s)}{V_{\text{I}}(s^{\prime})} + \dots.
\end{equation}
Higher-order terms involve increasingly nested commutators computed at different times. However, due to the properties of the $\text{su}(2)$ Lie algebra, $\Omega(t)$ can be written in terms of SU(2) generators as it is done in Eqs.~\eqref{eq:HamSU2},~\eqref{eq:InvSU2}. For our purposes, it is enough to truncate the series $\Omega(t)$ to the first order. Moreover, from Eq.\eqref{eq:auxFIR}, the interaction operator is easily written in the basis of the dynamical invariant, i.e., $V_{\text{I}}(t)=-(\delta/2)e^{-i(\beta(t)-\beta(t_{0}))}f_{\text{in}}(t)\ketbra{\phi_{+}}{\phi_{-}} +\text{h.c.}$, so that we can write
\begin{equation}\label{eq:magnu3}
U_{\text{I}}(t,t_{0})\simeq e^{-i\mathcal{M}(t)}, \mathcal{M}(t)=\abs{z(t)}e^{i\gamma(t)} \ketbra{\phi_{+}}{\phi_{-}} + \text{h.c.},
\end{equation}
where $\abs{z(t)}=(\delta/2)\sqrt{\mathcal{A}^2(t) +\mathcal{\tilde{A}}^2(t)},\gamma(t)=\arctan[\mathcal{A}(t)/\mathcal{\tilde{A}}(t)],\mathcal{A}(t)=\int_{t_{0}}^{t}\mathrm{d}s f_{\text{in}}(s)\cos\beta(s),\mathcal{\tilde{A}}(t)=\int_{t_{0}}^{t}\mathrm{d}s f_{\text{in}}(s)\sin\beta(s)$. Combining Eq.~\eqref{eq:magnu3}, Eq.~\eqref{eq:bounddef} and writing the initial state  in the invariant eigenbasis as explained in Sec.~\ref{sec:FIR}, it is thus easy to get back to the Schr\"odinger picture and write the state at the final time as follows
\begin{equation}\label{eq:magnu4}
\ket{\psi(t_{\text{f}})}=\frac{1}{\sqrt{2}}(\cos\abs{z(t_{\text{f}})} + ie^{i\gamma(t_{\text{f}})}\sin\abs{z(t_{\text{f}})})\ket{0}-\frac{1}{\sqrt{2}}(\cos\abs{z(t_{\text{f}})} + ie^{-i\gamma(t_{\text{f}})}\sin\abs{z(t_{\text{f}})})\ket{1}.
\end{equation}
From Eq.~\eqref{eq:magnu4}, the expectation values of $\sigma_z$ at final time reads
\begin{equation}\label{eq:magnu5}
\ev{\sigma_{z}(t_{\text{f}})} - 1= -(1-\cos(2\gamma(t_{\text{f}})))\sin^2\abs{z(t_{\text{f}})}.
\end{equation}
It is worth noticing that Eq.~\eqref{eq:magnu5} reduces to Eq.~\eqref{eq:pert20} in the limit $\mathcal{\tilde{A}}(t_{\text{f}})\to 0 $ and for sufficiently small amplitude $\delta$.
\section{Experimental details}\label{sec:expdetail}

\subsection{Experimental Setup}\label{sec:setup}
\textbf{Sample Preparation} \\
Experiments utilized a QZabre diamond substrate featuring engineered nanopillars to enhance photon extraction efficiency from individual Nitrogen-Vacancy (NV) centers. Individual pillars were selected based on optimal fluorescence intensity and isolation, identified via fluorescence microscopy.

\vspace{0.5em}

\textbf{Optical Setup} \\
An Olympus LMPLFLN 100X objective lens with high numerical aperture was employed to focus excitation light and collect fluorescence from NV centers. Optical initialization and readout of NV spin states were performed using a continuous-wave (CW) 520 nm laser diode (LABS Electronics), precisely focused onto individual nanopillars for targeted excitation.

\vspace{0.5em}

\textbf{Fluorescence Detection} \\
Fluorescence emission from the NV centers was collected through the same objective lens and filtered to remove residual excitation photons using a dichroic mirror. This mirror selectively transmitted red photons (\(\sim 630\,\text{nm}\)) to a single-photon counting module (SPCM-AQRH-13, Excelitas). Time-tagged single-photon counting provided temporal resolution for detailed fluorescence analysis. The NV center position was precisely determined by adjusting the diamond stage until maximum fluorescence was observed.

\vspace{0.5em}

\textbf{Magnetic Field and Microwave Control} \\
A cylindrical neodymium magnet, mounted on a motorized three-dimensional stage, applied a magnetic field to split the NV ground state spin levels (\(\ket{\pm 1_g}\)) via the Zeeman effect. Optimal Zeeman splitting was achieved through precise magnet positioning relative to the NV center.

Microwave signals required for spin manipulation were generated using an RF signal generator (SGS100A, Rohde \& Schwarz), amplified by a broadband RF amplifier (Minicircuits ZHL-5W-63-S+), and delivered to the NV center through a precisely positioned \(50\,\mu\text{m}\) diameter bond wire.

\vspace{0.5em}

\textbf{Spin Initialization and Measurement} \\
The NV center was initialized into its ground state \(\ket{0_g}\) using a CW 520 nm laser pulse. NV centers in \(\ket{0_g}\) transitioned to the excited state \(\ket{0_e}\), followed by decay back to \(\ket{0_g}\), emitting red photons. Centers in the \(\ket{\pm 1_g}\) states transitioned to \(\ket{\pm 1_e}\), subsequently decaying either radiatively (70\%) or non-radiatively through a dark state (30\%). The fluorescence intensity difference between states allowed for spin state determination through repeated measurements.

\vspace{0.5em}

\textbf{Rabi Oscillations and $\hat{\sigma}_x$ Calibration} \\
Optically Detected Magnetic Resonance (ODMR) spectroscopy identified the resonant transition frequency between the \(\ket{0_g}\) and \(\ket{-1_g}\) states, evident by a reduction in fluorescence intensity. The NV resonance frequency is related to the applied magnetic field \(B_z\) by:
\[
\omega_{NV} = D - \gamma_e B_z,
\]
where \(D = 2.87\,\text{GHz}\) is the zero-field splitting and \(\gamma_e = 2.8\,\text{MHz/G}\) is the electron gyromagnetic ratio.

Microwave pulses at the resonant frequency were applied for varying durations (0 to \(1\,\mu\text{s}\)), inducing coherent oscillations (Rabi oscillations) between spin states. Measured data were fitted to a decaying sinusoid to extract the Rabi frequency. The correlation between microwave amplitude (OPX voltage output) and the measured Rabi frequency was used to calibrate spin rotations along the \(\sigma_x\) axis.

\subsection{QIF Experimental Implementation}
\label{sec:I_filter_implement}
Equation~\eqref{eq:auxFIR} defines an exact relation between the invariant auxiliary field and the impulse response function. In practice, this relation can be simplified by directly choosing the auxiliary field in the form
\begin{equation}
\beta(t) = -\frac{\pi}{2} + \mathcal{H}(t),
\end{equation}
which also satisfies the invariant condition. This choice yields a control Hamiltonian that implements the desired filtered dynamics, allowing precise control over the transmitted signal's frequencies, phase, and amplitude.

\subsubsection{Construction of the Band-Pass 
Filter for Experimental Implementation}
\begin{figure}
\centering
\includegraphics[scale=0.3]{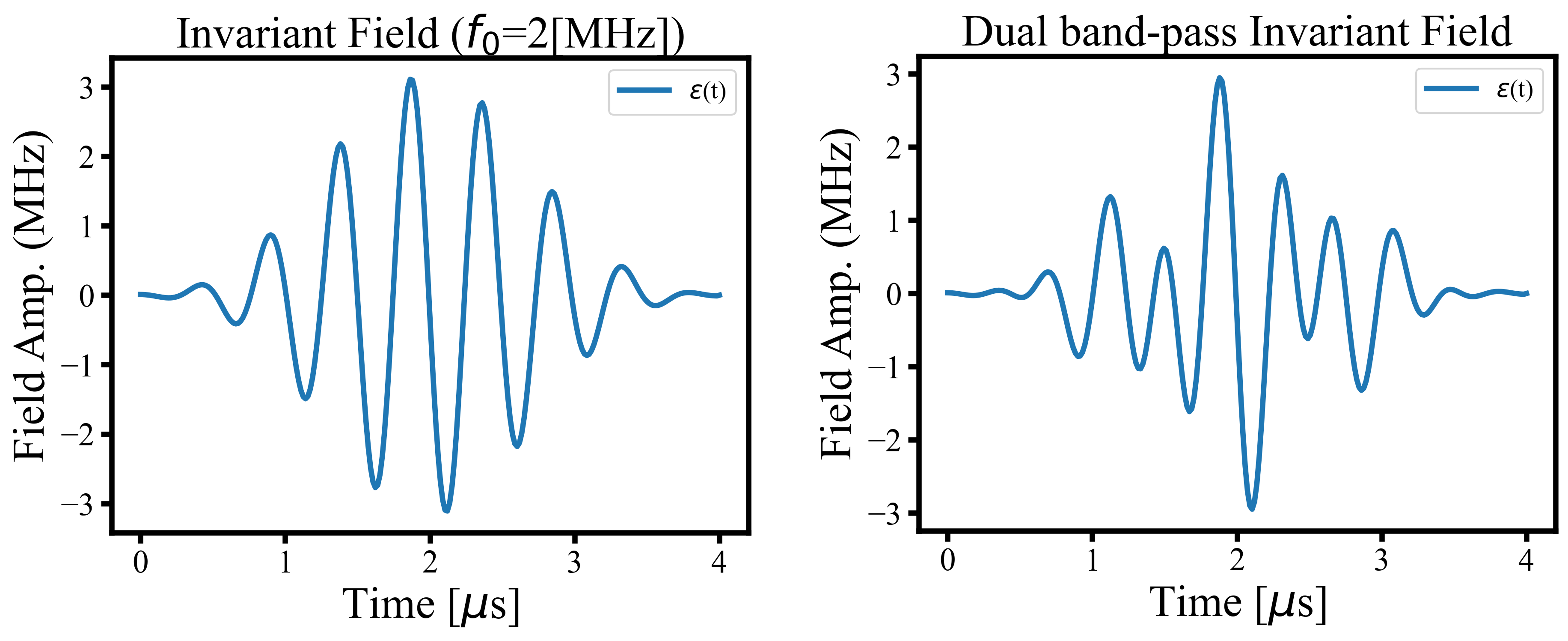}
\caption{Example of the computed control field $\varepsilon(t)$ using QIF protocol.}
\label{fig:control_fields}
\end{figure}

To realize the Quantum Invariant Filter (QIF) experimentally, we construct custom-designed band-pass filters that define the desired system response in the frequency domain. This is accomplished by embedding a classical Finite Impulse Response (FIR) filter into the time-dependent control Hamiltonian via the invariant formalism.

We begin by generating a low-pass FIR filter using Python’s \texttt{SciPy} signal processing library. The impulse response of this filter, denoted $\theta[n]$, is then mapped onto a smooth continuous function $\theta(t)$ to ensure compatibility with the invariant control formalism. This smoothing step is crucial: since the control fields depend on the time derivative of the auxiliary field, continuity and differentiability of $\theta(t)$ are required to avoid unphysical discontinuities in the control Hamiltonian.
To construct a band-pass filter centered at a desired frequency $f_0$, we multiply the continuous low-pass impulse response $\theta(t)$ by a cosine carrier:
\begin{equation}\label{eq:band_pass}
\mathcal{H}(t) = \theta(t) \cos\left(2\pi f_0 t_{\rm sym} + \phi\right),
\end{equation}
where $t \in [0, t_{\rm f}]$, $f_0$ is the central frequency of the filter, and $\phi$ is a tunable phase offset. This modulation shifts the filter to be centered around $f_0$ and simultaneously introduces phase sensitivity. The resulting $\mathcal{H}(t)$ serves as the impulse response function of the QIF protocol and is used to construct the auxiliary field $\beta(t)$ in the invariant formalism and consequently the control field $\varepsilon(t)$. Examples of such control fields are shown in Fig.~\ref{fig:control_fields}. 

In the experiment, this method provides a practical and flexible way to explore a wide range of filter frequencies and phases. Since the filter is defined analytically, scanning over different values of $f_0$ and $\phi$ can be performed efficiently without the need to redesign the full control protocol each time. This approach is particularly useful in measurements such as those shown in Fig.~2 and Fig.~3 of the main text, where phase resolution and frequency selectivity are demonstrated.
The construction of each filter requires setting the following key parameters:
\begin{itemize}
    \item $f_0$: Central frequency of the band-pass filter (in MHz)
    \item $t_{\rm f}$: Total duration of the control pulse sequence (in $\mu$s)
    \item \texttt{cutoff}: Bandwidth-defining cutoff frequency of the base FIR low-pass filter
    \item $F_s$: Sampling rate used in the digital filter design (samples per $\mu$s)
\end{itemize}

\begin{figure}[H]
\centering
\includegraphics[width=\textwidth]{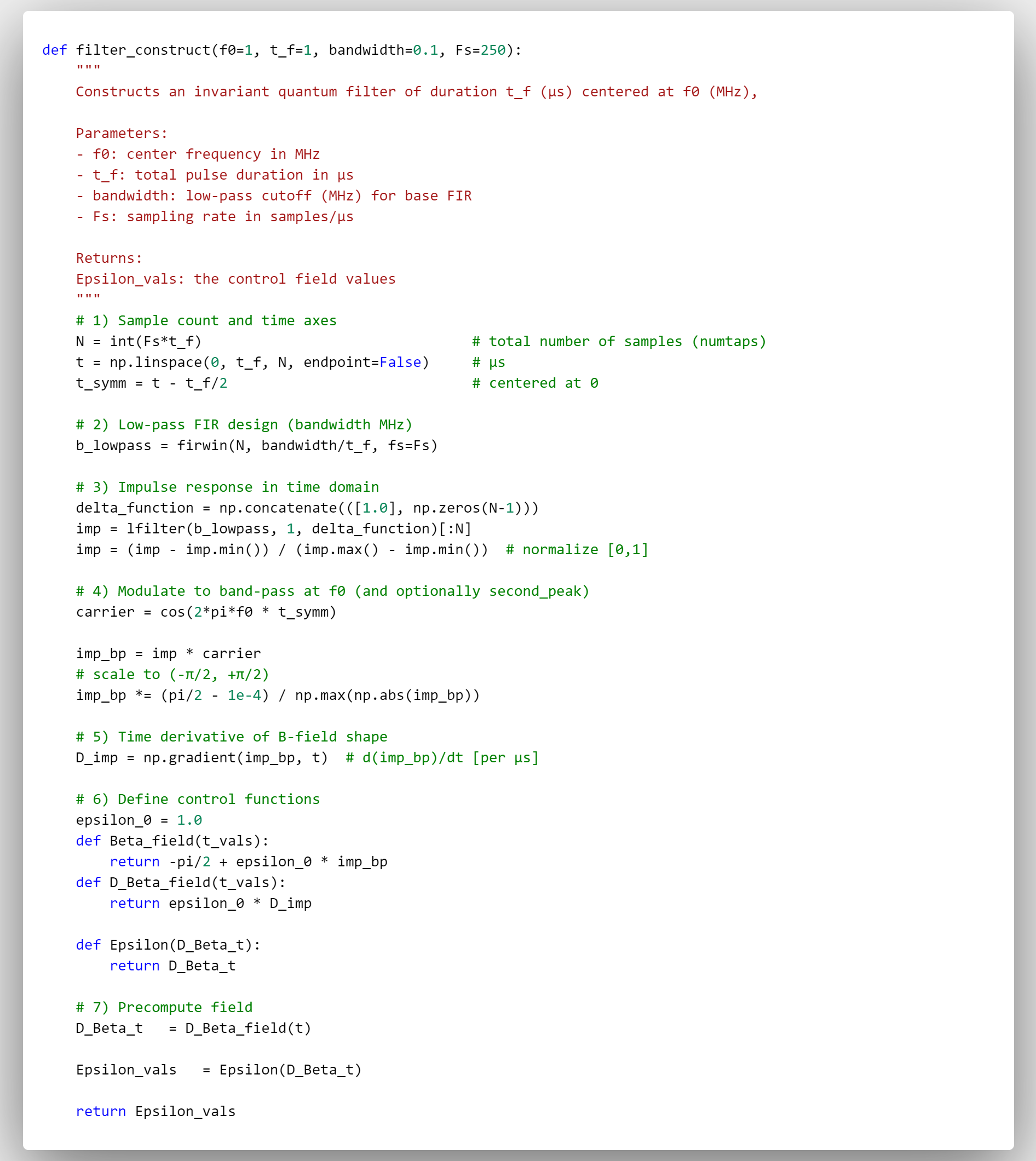}
\caption{Example code snippet generating the filter, the auxiliary fields, and the control field according to the QIF method.}
\label{fig:code-example}
\end{figure}

The symmetry of the filter is enforced by defining a symmetric time variable $t_{\text{sym}} = t - t_{\rm f}/2$, such that $\mathcal{H}(t)$ is centered in time. Boundary conditions are naturally satisfied by the FIR design, ensuring that $\mathcal{H}(0) = \mathcal{H}(t_{\rm f}) = 0$, as required by the invariant control framework.
This filtering procedure provides an efficient and experimentally viable route to realizing custom quantum filters with tailored spectral and phase characteristics. An example of the code is provided in Fig.~\ref{fig:code-example}.

\subsubsection{Experimental Characterization of Frequency and Phase Response}

Following the construction of the band-pass kernel $\mathcal{H}(t)$ described in the previous section, we now demonstrate experimentally that the QIF protocol faithfully reproduces its designed spectral and phase properties. The measured system response, quantified by the second-order amplitude $\mathcal{A}^2(t_{\rm f})$ (see Eq.~(\ref{eq:pert20})), is used as a direct probe of the filter characteristics.

\textbf{Frequency selectivity via cosine-modulated signal.}
To evaluate the frequency-domain response, we apply an external signal of the form
\begin{equation}
f_{\mathrm{in}}(t) = \delta \cos(2\pi f\, t_{\text{sym}}),
\end{equation}
and scan the frequency $f$ across a broad range. The resulting response amplitude is computed via the convolution Eq.~(\ref{eq:A}).
Due to the symmetry and compact support of the FIR-based envelope $\theta(t)$, this integral reduces to the cosine transform of $\mathcal{H}(t_{\text{sym}})$. When the probe frequency matches the filter frequency ($f = f_0$), constructive interference leads to a maximal response. For detuned frequencies, the integrand oscillates rapidly and the response is suppressed. As a result, the measured quantity $\mathcal{A}^2(t_{\rm f})$ reproduces the squared magnitude of the filter’s transfer function:
\begin{equation}
\mathcal{A}^2(t_{\rm f}) \propto \left|\mathcal{H}(f)\right|^2,
\end{equation}
where $\mathcal{H}(f)\equiv \mathcal{F}[\mathcal{H}(t)]$ denotes the Fourier transform of $\mathcal{H}(t)$. 

It is worth noting that when $\mathcal{H}(t)$ is a real and symmetric function (by construction $\mathcal{H}(t_{\text{sym}}) = \mathcal{H}(-t_{\text{sym}})$), the magnitude of its cosine transform is exactly equal to the magnitude of its Fourier transform:
\begin{equation}
\left| \int_{-\infty}^\infty \mathcal{H}(t)\, e^{-i2\pi f t}\, dt \right|
=
\left| \int_{-\infty}^\infty \mathcal{H}(t)\, \cos(2\pi f t)\, dt \right|.
\end{equation}
This identity ensures that the experimentally accessible cosine projection used in our measurements faithfully reproduces the absolute spectrum of the designed filter. 

\textbf{Phase sensitivity via sine-modulated signal.}
To characterize the phase resolution of the QIF, we apply an input signal at the filter frequency $f = f_0$, but modulated with a sine function:
\begin{equation}
f_{\mathrm{in}}(t) = \delta \sin(2\pi f_0\, t_{\text{sym}}).
\end{equation}
This signal is orthogonal to the filter carrier at $\phi = 0,\pm\pi$, enabling a controlled study of the system’s response as a function of the filter phase $\phi$. Substituting into the convolution expression yields
\begin{equation}
\mathcal{A}(t_{\rm f}) = \delta \int_{-t_{\rm f}/2}^{t_{\rm f}/2} \theta(t_{\text{sym}})\,
\cos(2\pi f_0 t_{\text{sym}} + \phi)\, \sin(2\pi f_0 t_{\text{sym}})\,dt_{\text{sym}}.
\end{equation}
Using the identity 
$\cos(a + \phi)\sin a = \frac{1}{2}[\sin(\phi) + \sin(2a + \phi)]$, 
and noting that the second term averages to zero due to rapid oscillations, we obtain
\begin{equation}
\mathcal{A}(t_{\rm f}) \propto \sin(\phi) \int_{-t_{\rm f}/2}^{t_{\rm f}/2} \theta(t_{\text{sym}})\, dt_{\text{sym}},
\quad \Rightarrow \quad
\mathcal{A}^2(t_{\rm f}) \propto \sin^2(\phi).
\end{equation}

This result confirms that the QIF is not only frequency-selective, but also phase-sensitive. The system response is maximized when the signal is in quadrature with the filter ($\phi = \pm \pi/2$) and vanishes when they are in phase or anti-phase ($\phi = 0, \pm \pi$). This behavior, shown in Fig.~3(a–c) of the main text, demonstrates coherent phase discrimination, an essential feature of lock-in-style detection, now realized in a fully programmable quantum control protocol.

These experimental results validate that the QIF protocol implements a customizable band-pass filter whose spectral profile and phase sensitivity can be engineered independently through the construction of $\mathcal{H}(t)$. This capability enables flexible, high-fidelity filtering of coherent quantum signals with tunable bandwidth, frequency, and phase selectivity, all defined directly in the frequency domain and mapped to control fields through the invariant formalism.

\textbf{Multi-mode selectivity via cosine-modulated signals.}
The QIF framework further enables the design and implementation of multi-mode filters, allowing simultaneous selectivity at multiple target frequencies. This is achieved by combining multiple frequency components within the filter kernel. In the experiment shown in Fig.~3(e–f) of the main text, we implemented a dual-band filter by constructing the impulse response as
\begin{equation}
\mathcal{H}(t) = \frac{1}{2}\theta(t)\left[\cos(2\pi f_1\, t_{\text{sym}}) + \cos(2\pi f_2\, t_{\text{sym}})\right],
\end{equation}
where $f_1$ and $f_2$ denote the two central frequencies of interest. The low-pass envelope $\theta(t)$ ensures bandwidth control and time-domain boundary compliance, while the superposition of cosine carriers shifts the filter's response to both $f_1$ and $f_2$. As a result, the corresponding transfer function exhibits well-defined peaks around these two frequencies, confirming multi-mode selectivity.
This construction highlights the flexibility of the QIF approach: arbitrary classical filters, including those generated using standard signal-processing libraries, can be translated into valid control Hamiltonians for quantum systems.

\subsection{Field Generation Using the OPX}
To generate these fields, we used the OPX quantum controller (Quantum Machines), which offers 4 ns resolution in arbitrary waveform generation (250 samples per $\mu\text{s}$). The total number of RF pulse samples was computed as:
\begin{equation*}
\text{Number-of-Samples} = 250 \times t_{\rm f},
\end{equation*}

where $t_{\rm f}$ is in $\mu\text{s}$.
The OPX imposes a $2^{16}$ sample limit per pulse ($\sim 65536$ samples, corresponding to about $260\,\mu\text{s}$). For longer pulse sequences, the resolution was reduced, sampling every 16 ns or 32 ns instead of every 4 ns. This reduction was only necessary for high-$f_0$ filters, where the field oscillated multiple times within a short duration.


\textbf{Polarization:}
All measurements in this sequence were carried out under an external magnetic field of approximately 670 Gauss, which produced a 1 GHz resonant transition and effectively polarized the $N^{15}$ nuclear spin into a single state.

\textbf{CPMG Sequence Figure}
The protocol begins with $\pi$ pulse calibration, where Rabi oscillations are fitted to determine the precise duration required for a $\pi$ rotation. This duration is then rounded to the nearest 4 ns increment, and the RF amplitude is adjusted accordingly. To fine tune the amplitude, a CPMG n sequence is executed without any applied signal; the sharpness of the echo peak, which scales with n, indicates the optimal amplitude. For two dimensional CPMG data acquisition, the number of $\pi$ pulses (n) is varied, and for each n the signal frequency is swept across the desired range. One million repetitions are performed at each frequency, and the results are averaged. Between experiments at different n values, an auto track routine re-centers the NV center on its maximum fluorescence, and the Rabi calibration is repeated to correct for any frequency shifts caused by slight wire bond movements .

\textbf{Filter Sequence Figure}
Each filter in the figure has a fixed duration of 4 $\mu$s. For every chosen center frequency $f_0$, the corresponding time dependent control field $\epsilon$(t) is reverse engineered. During data acquisition, the filter field is applied along the $\sigma_x$ axis while a test signal is simultaneously applied along $\sigma_z$. For each signal frequency, the NV state is measured and then reinitialized; one million cycles are averaged to yield the response. To quantify filter performance, each row of the figure is collected twice: once with the NV initialized in $\ket{0}$ and once in $\ket{1}$. The difference between these two data sets defines the filter contrast .

\textbf{Amplitude Detection Figure}
In this experiment, the filter design remains fixed at 4 $\mu$s duration with a center frequency $f_0$ of 2 MHz and a test signal also at 2 MHz. The variable parameter is the amplitude of the applied test signal. During each measurement, the filter field is applied on $\sigma_x$, followed by the test signal on $\sigma_z$. For each amplitude setting, the NV is measured and reinitialized, and one million cycles are averaged. Contrast is determined by collecting each point twice—once with the NV in $\ket{0}$ and once in $\ket{1}$—and taking their difference, which is then translated into the $<Z>$ component of the NV spin.

\textbf{Signal Phase Detection Figure}
The filter generation multiplies a low-pass FIR filter (designed using SciPy) by a $\cos(2\pi f_0 t)$ carrier, passing only signals in phase with the filter cosine. By introducing a phase shift $\phi$ into the carrier (multiplying the FIR response by $\cos(2\pi f_0 t + \phi)$), the phase of the incoming signal can be detected. In this sequence, the duration of the filter is fixed at 4~$\mu\text{s}$ and $f_0$ is set to 1.5~MHz. Each row of the figure corresponds to a specific filter phase while sweeping the signal frequency. For data acquisition, the filter field is applied on $\sigma_x$ and the test signal on $\sigma_z$; one million repetitions per frequency are averaged after measuring and re-initializing the NV. Contrast is again obtained by differencing measurements from NV initialized in $\ket{0}$ versus $\ket{1}$, and translating into the $\langle Z \rangle$ component.

\textbf{$T_1$ and $T_2$ Figure}
Measurements for $T_1$ and $T_2$ were performed on a different NV center which haven’t been polarized. For the filter data, the filter passing frequency is maintained constant relative to the sequence duration by scaling $f_0$ in proportion to each measured duration $t_i$. In the CPMG 8 and CPMG 16 experiments, the time interval between $\pi$ pulses is swept. For the $T_1$ measurement, the NV is initialized only in $\ket{0}$, allowed to evolve for a variable delay $T_i$ (in $\mu$s), and then measured. In every case—filter, CPMG, and $T_1$—the control field (or $\pi$ pulse sequence for CPMG) is applied, the NV is measured and reinitialized for each duration, and one million cycles are averaged. Contrast is determined by collecting data with the NV initialized in $\ket{0}$ and $\ket{1}$, with the difference yielding the final $<Z>$ component.

\end{document}